\documentclass[twoside,twocolumn,9pt]{article}
\usepackage{extsizes}
\usepackage[super,sort&compress,comma]{natbib}
\usepackage[left=1.5cm, right=1.5cm, top=1.785cm, bottom=2.0cm]{geometry}
\usepackage{balance}
\usepackage{mathptmx}
\usepackage{sectsty}
\usepackage{graphicx}
\usepackage{lastpage}
\usepackage{mathtools}
\usepackage{cuted}
\usepackage[format=plain,justification=justified,singlelinecheck=false,font={stretch=1.125,small,sf},labelfont=bf,labelsep=space]{caption}
\usepackage{float}
\usepackage{fancyhdr}
\usepackage{fnpos}
\usepackage[english]{babel}
\addto{\captionsenglish}{%
  \renewcommand{\refname}{Notes and references}
}
\usepackage{array}
\usepackage{droidsans}
\usepackage{charter}
\usepackage[T1]{fontenc}
\usepackage[usenames,dvipsnames]{xcolor}
\usepackage{setspace}
\usepackage[compact]{titlesec}
\usepackage{hyperref}

\usepackage{epstopdf}%This line makes .eps figures into .pdf - please comment out if not required.
\usepackage{extarrows}
\definecolor{cream}{RGB}{222,217,201}
\newcommand{\ib}[1]{{\color{black}#1}}

\newcommand{\iv}{$I$\textendash$V$}

\newcommand{\figname}{Fig.~}

\newcommand{\figsname}{Figs.~}

\newcommand{\gl}{eqn}

\newcommand{\Gl}{Eqn}

\newcommand{\alt}{\raisebox{-0.3ex}{$\stackrel{<}{\sim}$}}

\usepackage[version=4]{mhchem} % Formula subscripts using \ce{}
\usepackage{threeparttable}
\DeclareGraphicsExtensions{.jpg, .pdf, .png}
\graphicspath{{/media/ioan/Elements/gt/5by3/data/}{/media/ioan/Elements/gt/5gby3/}{.}}

\begin{document}

\pagestyle{fancy}
\thispagestyle{plain}
\fancypagestyle{plain}{
\renewcommand{\headrulewidth}{0pt}
}
\makeFNbottom
\makeatletter
\renewcommand\LARGE{\@setfontsize\LARGE{15pt}{17}}
\renewcommand\Large{\@setfontsize\Large{12pt}{14}}
\renewcommand\large{\@setfontsize\large{10pt}{12}}
\renewcommand\footnotesize{\@setfontsize\footnotesize{7pt}{10}}
\makeatother

\renewcommand{\thefootnote}{\fnsymbol{footnote}}
\renewcommand\footnoterule{\vspace*{1pt}%
\color{cream}\hrule width 3.5in height 0.4pt \color{black}\vspace*{5pt}}
\setcounter{secnumdepth}{5}

\makeatletter
\renewcommand\@biblabel[1]{#1}
\renewcommand\@makefntext[1]%
{\noindent\makebox[0pt][r]{\@thefnmark\,}#1}
\makeatother
\renewcommand{\figurename}{\small{Fig.}~}
\sectionfont{\sffamily\Large}
\subsectionfont{\normalsize}
\subsubsectionfont{\bf}
\setstretch{1.125} %In particular, please do not alter this line.
\setlength{\skip\footins}{0.8cm}
\setlength{\footnotesep}{0.25cm}
\setlength{\jot}{10pt}
\titlespacing*{\section}{0pt}{4pt}{4pt}
\titlespacing*{\subsection}{0pt}{15pt}{1pt}
\fancyfoot{}
\fancyfoot[RO]{\footnotesize{\sffamily{1--\pageref{LastPage} ~\textbar  \hspace{2pt}\thepage}}}
\fancyfoot[LE]{\footnotesize{\sffamily{\thepage~\textbar\hspace{4.65cm} 1--\pageref{LastPage}}}}
\fancyhead{}
\renewcommand{\headrulewidth}{0pt}
\renewcommand{\footrulewidth}{0pt}
\setlength{\arrayrulewidth}{1pt}
\setlength{\columnsep}{6.5mm}
\setlength\bibsep{1pt}
\makeatletter
\newlength{\figrulesep}
\setlength{\figrulesep}{0.5\textfloatsep}

\newcommand{\topfigrule}{\vspace*{-1pt}%
\noindent{\color{cream}\rule[-\figrulesep]{\columnwidth}{1.5pt}} }

\newcommand{\botfigrule}{\vspace*{-2pt}%
\noindent{\color{cream}\rule[\figrulesep]{\columnwidth}{1.5pt}} }

\newcommand{\dblfigrule}{\vspace*{-1pt}%
\noindent{\color{cream}\rule[-\figrulesep]{\textwidth}{1.5pt}} }

\makeatother
\twocolumn[
  \begin{@twocolumnfalse}
\vspace{1em}
\sffamily

\begin{tabular}{m{4.5cm} p{13.5cm} }

  & \noindent\LARGE{\textbf{Gaining insight into molecular tunnel junctions with a pocket calculator without
      $I$\textendash$V$ data fitting. Five-thirds protocol}} \\%Article title goes here instead of the text "This is the title"
\vspace{0.3cm} & \vspace{0.3cm} \\

 & \noindent\large{Ioan B\^aldea\textit{$^{a}$}} \\%Author names go here instead of "Full name", etc.

& \noindent\normalsize{The protocol put forward in the present paper is an attempt to meet the experimentalists' legitimate desire
  of reliably and easily extracting microscopic parameters from current-voltage measurements on molecular junctions.
  It applies to junctions
  wherein charge transport dominated by a single level (molecular orbital, MO) occurs via off-resonant tunneling. The recipe is simple.
  The measured \ib{current-voltage} curve \ib{$I = I(V)$}
  should be recast as a curve of $V^{5/3}/I$ versus $V$. \ib{This curve exhibits two maxima:
    one at positive bias ($V = V_{p+}$), another at negative bias ($V = V_{p-}$).}
  The values $V_{p +} > 0$ and $V_{p -} < 0$ at the two peaks
  of the curve for $V^{5/3}/I$ at positive and negative bias and the corresponding values \ib{$I_{p +} = I(V_{p+}) > 0$} and
  \ib{$I_{p -} = I(V_{p-}) < 0$} of the current is all information needed \ib{as input}.
  The arithmetic average of $V_{p +}$ and $\vert  V_{p -}\vert$ in volt provides the value in electronvolt of the MO energy offset $\varepsilon_0 = E_{MO} - E_F$ relative to the electrode Fermi level ($\vert \varepsilon_0\vert = e (V_{p +} + \vert V_{p -}\vert )/2$). The value of the (Stark) strength of the bias-driven MO shift is obtained as $\gamma = (4/5) (V_{p +} - \vert V_{p -} \vert) / (V_{p +} + \vert V_{p -} \vert) $. Even the low-bias conductance estimate, $ G = (3/8) (I_{p +} / V_{p +} + I_{p -} / V_{p -})$, can be a preferable alternative to that deduced from fitting the $I$\textendash$V$ slope in situations of noisy curves at low bias. To demonstrate the reliability and the generality of this ``five-thirds'' protocol, I illustrate its wide applicability for molecular tunnel junctions fabricated using
  \ib{metallic and nonmetallic electrodes, molecular species possessing localized $\sigma$ and delocalized $\pi$ electrons, and} various techniques (mechanically controlled break junctions, STM break junctions, conducting probe AFM junctions, and large area junctions).} \\

\end{tabular}

 \end{@twocolumnfalse} \vspace{0.6cm}
  ]
\renewcommand*\rmdefault{bch}\normalfont\upshape
\rmfamily
\section*{}
\vspace{-1cm}

\footnotetext{\textit{$^{a}$~Theoretical Chemistry, Heidelberg University, Im Neuenheimer Feld 229, D-69120 Heidelberg, Germany. Fax: +49 6221 545221; Tel: +49 6221 545219; E-mail: ioan.baldea@pci.uni-heidelberg.de}}

\section{Introduction}
\label{sec:intro}

Conventional semiconductor microelectronics has at its disposal a series of simplified equations
to easily gain insight into underlying physics.\cite{Sze:06,Thompson:14}
This is the case, e.g., of the familiar Shockley equation (``ideal diode law''),\cite{Shockley:49,Shockley:50,Sah:57,Moll:58}
expressing analytically the exponential dependence of the current $I$ on bias $V$ stemming from the microscopically built-in potential barrier
at a p-n junction.

In an attempt to establish the molecular structure-tunneling transport relationship,
starting from ideas put forward by Newns and Schmickler in conjunction with electro/chemisorption,\cite{Newns:69b,Anderson:61,Schmickler:86}
I have deduced, as a counterpart for molecular electronics, an appealingly simple formula for molecular junctions 
wherein thermal effects are ignored \cite{Baldea:2023a,Baldea:2024a}) and
the off-resonant tunneling current is dominated by a single level (molecular orbital MO).\cite{Baldea:2012a}

This off-resonant single level model (orSLM) expresses the current as a function of bias in terms of three
key electronic structure parameters:
the MO energy offset relative to electrodes' Fermi energy $\varepsilon_0 = E_{MO} - E_F$,
the average MO coupling $\Gamma = \sqrt{\Gamma_s \Gamma_t}$ to the two (generic substrate $s$ and tip $t$) electrodes,
and the bias-driven MO shift $\gamma$
\begin{equation}
  \label{eq-jIB}
  I \equiv I(V; \varepsilon_0, \gamma, G) = \frac{G \varepsilon_0^2 V}{\varepsilon_V^2 - (e V/2)^2};\
  \varepsilon_V = \varepsilon_0 + \gamma e V;\
  G = N G_0 \frac{\Gamma^2}{\varepsilon_0^2}
\end{equation}
Above, $G_0 = 2 e^2/h = 77.48\,\mu$S is the conductance quantum and $N$ is the number of molecules per junction.
Within \gl~(\ref{eq-jIB}), asymmetric {\iv} curves (current rectification)
correspond to junctions where the MO energy is shifted by an applied bias ($\gamma \neq 0$). {\iv} curves are symmetric
(no current rectification) in the absence of a bias-driven MO shift ($\gamma = 0$, $\varepsilon_V = \varepsilon_0$).
To be sure, aiming at describing charge transport by tunneling in situations where thermal effects are negligible,
\gl~(\ref{eq-jIB}) has limits of validity precisely formulated.\cite{Baldea:2023a,Baldea:2024a}
They should strictly be observed in specific applications to real molecular junctions.

Methodological advantages of the orSLM approach have been highlighted recently.\cite{Frisbie:23}
One particular aspect worth mentioning is the excellent agreement between the MO offset extracted from transport data
and the MO offset obtained from
a completely different method, namely independent ultraviolet photoelectron spectroscopy (UPS) data.
This finding is all the more important, since it has been reported for completely different homologous molecular series.\cite{Baldea:2019d,Baldea:2019h,Frisbie:21a}
This agreement represents perhaps the strongest support that validates the analysis based on the orSLM.

Fitting measured $I$\textendash$V$ data to \gl~(\ref{eq-jIB}) and extracting best fitting parameters $\varepsilon_0$, $\Gamma$
(or the low bias conductance $G$, to which I will loosely refer as a ``microscopic'' parameter in order to obviate lengthy discussions for junctions with $ N > 1$),
and $\gamma$ should pose no special problem. This is confirmed by numerous applications of the orSLM approach
by many independent groups,
\cite{Vuillaume:12a,Vuillaume:12c,Fracasso:13,Tao:13,Hou:13,Yu:15,Ho:15,Yutaka:15,Chiechi:15a,Guo:16a,Guo:16b,Lee:16,Lee:16a,Yu:16b,Yu:16c,Yu:16d,Lenfant:17,Yu:18,Song:18a,Cuevas:19a,Gu:21,Chiechi:21,Song:22a,Chiechi:22,Jang:23}
which succeeded in correctly reproducing $I$\textendash$V$ curves measured for molecular junctions fabricated using various techniques.

Notwithstanding the aforementioned, publications also exist wherein, unfortunately,
model parameter values have been incorrectly determined data fitting to \gl~(\ref{eq-jIB}).
Because emphasis in this paper is on how to easily and correctly estimate microscopic parameters by means of the orSLM,
I intentionally refrain from citing misapplications of this approach. Still, I want to mention a frequent error that can be
immediately identified.
Those publications report MO offsets \(\vert\varepsilon_0\vert( < e V/2)\) incompatible
with the bias range \((-V, +V)\) used for data fitting. 
This can be easily seen by inspecting \gl~(\ref{eq-jIB}), wherein the denominator becomes negative at too higher biases,
beyond the scope of this model.
This corresponds to a completely nonphysical situation wherein the current and bias have opposite signs.
In fact, as reiterated again and again,\cite{Baldea:2023a,Baldea:2024a} a necessary (``off-resonance'') condition for
\gl~(\ref{eq-jIB}) to apply is that of sufficiently lower biases
(usually \( e \vert V\vert\ \alt 1.4 \vert\varepsilon_0\vert\)).

Below, I will show that and how the microscopic parameters $\varepsilon_0$, $\gamma$ and $G$
can be directly estimated from appropriately recasting the measured  $I$\textendash$V$ curves
obviating the usage of \gl~(\ref{eq-jIB}) with adjustable model parameters which could lead to the unpleasant
situations referred to in the preceding paragraph.

\ib{Before proceeding, let me emphasize what is already expressed by the title of this paper.
  My present aim is to demonstrated that,
  provided that conditions of validity clearly stated are fulfilled (see refs.~\citenum{Baldea:2023a}
  and \citenum{Baldea:2024c} and \figname\ref{fig:background} below),
  the orSLM allows to estimate microscopic parameters characterizing real molecular tunnel junctions
  obviating {\iv} data fitting. For this reason, a comparison with other data fitting approaches from the literature
  \cite{Schmickler:97,Tao:97,Reed:03,Wandlowski:08,Nijhuis:16a,Cornil:20,Song:22b} would be misplaced and will not be attempted.}
 
\section{Basic working equations}
\label{sec:eqn} 
Aiming at providing a theoretical basis and generalizing the transition voltage spectroscopy (TVS) approach
proposed by Frisbie et al,\cite{Beebe:06}
I showed that the parameters $\varepsilon_0$ and $\gamma$ can be estimated from the so-called
transition voltages $V_{t\pm}$ \cite{Baldea:2012a}
\begin{subequations}
  \label{eq-tvs}
\begin{eqnarray}
  \label{eq-vt-tvs}
  \left\vert\varepsilon_0\right\vert & = & 2
  \frac{e V_{t +} \left\vert V_{t -}\right\vert}{\sqrt{V_{t +}^2 + 10 V_{\kappa +}\left\vert V_{\kappa -}\right\vert/3 + V_{\kappa -}^2}} \\
  \label{eq-gamma-tvs}
  \gamma & = & \frac{\mbox{sign}\, \varepsilon_0}{2}
  \frac{V_{t +} - \left\vert V_{t -}\right\vert}{\sqrt{V_{t +}^2 + 10 V_{t +}\left\vert V_{t -}\right\vert / 3 + V_{t -}^2}}  
\end{eqnarray}
\end{subequations}
$V_{t +}(>0)$ and $V_{t -}(<0)$ can be defined as the positive and negative values of the bias where
$\ln \vert I/V^2 \vert $ is minimum\cite{Beebe:06} or
$V^2 /\vert I\vert$ is maximum,\cite{Baldea:2015a,Baldea:2015b} or, mathematically equivalently,
where the differential conductance is two times larger than the nominal (pseudo-ohmic) conductance \cite{Baldea:2012e}
\begin{equation}
  \label{eq-def-vt}
  \left . \frac{V^2}{\vert I\vert} \right\vert_{V = V_t} = \mbox{max} \Leftrightarrow
  \left . \frac{\partial I}{\partial V}\right\vert_{V = V_t} = 2 \left . \frac{I}{V}\right\vert_{V=V_t}
\end{equation}

\ib{\Gl~(\ref{eq-def-vt}) can be easily deduced:\cite{Baldea:2012a} one should plug the expression for the current (\gl~(\ref{eq-jIB})
into \gl~(\ref{eq-def-vt}) and solve the ensuing quadratic equation.}
  
\Gl~(\ref{eq-def-vt}) is a particular case ($\kappa = 2$) of a more general condition
\begin{equation}
  \label{eq-def-v-kappa}
  \left . \frac{\vert V\vert ^\kappa}{\vert I\vert} \right\vert_{V = V_\kappa} = \mbox{max} \Leftrightarrow
  \left . \frac{\partial I}{\partial V}\right\vert_{V = V_\kappa} = \kappa \left . \frac{I}{V}\right\vert_{V=V_\kappa}
\end{equation} 
for which the counterpart of the particular \gl~(\ref{eq-tvs}) can also be deduced analytically \cite{Baldea:2012a}
\begin{subequations}
  \label{eq-kappa}
  \begin{eqnarray}
  \label{eq-v-kappa}
  \left\vert\varepsilon_0\right\vert & = & \frac{\kappa ( \kappa + 1)}{\kappa^2 - 1}
  \frac{e V_{\kappa +} \left\vert V_{\kappa -}\right\vert}{\sqrt{V_{\kappa +}^2 + 2 \frac{\kappa^2 + 1}{\kappa^2 - 1} V_{\kappa +}\left\vert V_{\kappa -}\right\vert + V_{\kappa -}^2}} \\
\label{gamma-kappa}
  \gamma & = & \frac{\mbox{sign}\, \varepsilon_0}{2}
  \frac{V_{\kappa +} - \left\vert V_{\kappa -}\right\vert}{\sqrt{V_{\kappa +}^2 + 2 \frac{\kappa^2 + 1}{\kappa^2 - 1} V_{\kappa +}\left\vert V_{\kappa -}\right\vert + V_{\kappa -}^2}}  
\end{eqnarray}
\end{subequations} 

Above, $V_{\kappa +}$ and $V_{\kappa -}$
are the positive and negative biases at the peaks of the plot of $\vert V\vert^{\kappa}/\vert I \vert $
versus $V$.
\ib{\Gl~(\ref{eq-kappa}) can also be easily deduced:\cite{Baldea:2012a} one should plug the expression for the current (\gl~(\ref{eq-jIB})
into \gl~(\ref{eq-def-v-kappa}) and solve the quadratic equation thus obtained.}

As will be seen shortly below, the formulas for $\kappa = 5/3$ are particularly interesting
\begin{subequations}
\label{eq-5/3}
  \begin{eqnarray}
\label{eq-v-5/3}
  \left\vert\varepsilon_0\right\vert & = & \frac{5}{2}
  \frac{e V_{ {p}  +} \left\vert V_{ {p}  -}\right\vert}{\sqrt{V_{ {p}  +}^2 + \frac{17}{4} V_{ {p}  +}\left\vert V_{ {p}  -}\right\vert + V_{ {p}  -}^2}} \\
\label{eq-gamma-5/3}
  \gamma & = & \frac{\mbox{sign}\, \varepsilon_0}{2} \frac{V_{ {p}  +} - \left\vert V_{ {p}  -}\right\vert}{\sqrt{V_{ {p}  +}^2 + \frac{17}{4} V_{ {p}  +}\left\vert V_{ {p}  -}\right\vert + V_{ {p}  -}^2}}
\end{eqnarray}
In view of the special role played by the value $\kappa = 5/3$ anticipated above, I will write
$V_{{ p }\pm}$ instead of $ V_{{ 5/3 } \pm }$ to specify the location of the peaks of $\vert V\vert ^{5/3}/\vert I \vert$
\begin{equation}
  \frac{\vert V\vert^{5/3}}{\vert I\vert} = \max \Leftrightarrow V = V_{{ p }\pm} ( \equiv V_{ 5/3, \pm})
\end{equation}
\end{subequations}

In principle, the parameters $\varepsilon_0$ and $\gamma$ can be computed from $V_{ {p}  \pm}$ via \gl~(\ref{eq-5/3})
just as these parameters can be calculated from the transition voltages $V_{t\pm}$ via \gl~(\ref{eq-tvs})
(or, in general, from $V_{\kappa \pm}$ via \gl~(\ref{eq-kappa})).

The reason why $\kappa = 5/3$ is a special value becomes clear by considering the case of symmetric $I$\textendash$V$ curves
($\gamma = 0$, cf.~\gl~(\ref{eq-jIB})).
In such cases the peaks of $V^{5/3}/I$ are located symmetrically around origin
($V_{{ 5/3 } +} = \vert V_{{ 5/3 } - } \vert \equiv V{_{p}}$) and \gl~(\ref{eq-v-5/3}) reduces to
\begin{equation}
  \label{eq-v-p-symm}
  \gamma = 0 \rightarrow \left\vert\varepsilon_0\right\vert = e V_{p} 
\end{equation}
That is, \gl~(\ref{eq-v-p-symm}) allows the most straightforward determination of the MO energy offset
from the current-voltage measurements.
What one has to do in order to estimate the MO offset $\varepsilon_0$ of a junction
with symmetric $I$\textendash$V$ characteristic is merely to draw a 
plot of $V^{5/3}/I$ versus $V$. Expressed in electronvolt,
the value of the MO energy offset ($\varepsilon_0$) is equal to the magnitude in volt of
the bias ($V_{p} = V_{p+} = \vert V_{p-}\vert $)
at which the two symmetric peaks of the curve for $V^{5/3}/I$ are located.

To be sure, \gl~(\ref{eq-v-p-symm}) applies to molecular junctions having symmetric $I$\textendash$V$ characteristics
($I(-V) = - I(V)$) while real junctions often possess asymmetric characteristics ($I(-V) \neq - I(V)$).
Within the orSLM this asymmetry (current rectification) stems from a nonvanishing value $\gamma \neq 0$ (cf.~\gl~(\ref{eq-jIB})).

Because a nonvanishing $\gamma$ is directly related (cf.~\gl~(\ref{eq-gamma-5/3})) to an asymmetric location 
of the peaks of $V^{5/3}/I$ around origin ($V_{ {p} +} \neq - V_{ {p} -}$), it makes sense to define an average
peak voltage $V_{p}$ and to consider \ib{Taylor} series expansions of \gl~(\ref{eq-5/3})
in terms of the departure $\delta V_p$ of $V_{ {p} \pm}$ from the average $V_{p}$
\begin{subequations}
  \begin{equation}
    \label{eq-vp}
  V_{p} \equiv \frac{V_{ {p} +} + \vert V_{ {p} -}\vert}{2} = \frac{V_{ {p} +} - V_{ {p} -}}{2} ( > 0 )
\end{equation}
\begin{equation}
  V_{ {p} \pm} = \pm V_{p} + \delta V_{p}; \ 
\delta V_{p} = \frac{V_{ {p} +} - \vert V_{ {p} -} \vert}{2} \left(\stackrel{\leq}{>} 0\right)
\end{equation}
\end{subequations}
The lowest order expansion of \gl~(\ref{eq-5/3}) \ib{(i.e., neglecting all powers of $\delta V_p$ in the Taylor series)}
yields the following approximate expressions 
\begin{subequations}
\label{eq-approx}
  \begin{eqnarray}
\label{eq-e0-approx}
  \vert \varepsilon_0 \vert & \approx & \left\vert \varepsilon_{0}^{\mbox{\small approx}} \right\vert = e V_{p} = e \frac{V_{ {p} +} + \vert V_{ {p} -}\vert }{2}\\
\label{eq-gamma-approx}
  \gamma & \approx & \gamma^{\mbox{\small approx}} = \frac{2}{5} \frac{\delta V_{p}}{V_{p}} \mbox{sign}\, \varepsilon_0 
  = \frac{4}{5} \frac{V_{{p} +} - \vert V_{ {p} -}\vert}{V_{ {p} +} + \vert V_{ {p} -}\vert} \mbox{sign}\, \varepsilon_0 \\
\label{eq-G-approx}
  G & \approx & G^{\mbox{\small approx}} = \frac{3}{8}\left(
  \frac{I_{ {p} +}}{V_{ {p} +}} + \frac{I_{ {p} -}}{V_{ {p} -}}\right); \ I_{ {p} \pm } \equiv I\left(V_{ {p} \pm}\right)
\end{eqnarray}
\end{subequations}
\Gl~(\ref{eq-approx}) shows that all three parameters $\varepsilon_0$, $\gamma$, and $G$
that microscopically characterize a tunneling junction can be estimated from four
experimental quantities only, which can directly extracted from {\iv} measurements: the positive and negative bias $V_{ {p} \pm} $
where the peaks of the curve $\vert V\vert^{5/3} / \vert I\vert $ are located and the corresponding
currents $I_{ {p} \pm} $. 

Notice that in addition to $\varepsilon_0$, $\gamma$, \gl~(\ref{eq-G-approx}) presents an approximate estimate for
the low bias conductance $G$. It has been deduced by series expansion of $G$ expressed using \gl~(\ref{eq-jIB}).
Although $G$ is routinely determined by linear fitting of low $V$ data, the estimate via
\gl~(\ref{eq-G-approx}) may be preferable in cases of noisy data at low bias.
\ib{This may be a relevant aspect for reliably determining the tunneling attenuation $\beta$
  from conductances $G_n \propto \exp(-\beta n)$ of menbers of variable size $n$ of a homologous series.}  

\section{Accuracy of the lowest order approximation}

The smaller the asymmetry $\delta V_p $ (or, alternatively, the smaller the value or $\gamma$),
the better the lowest order approximation underlying \gl~(\ref{eq-approx}),
but the question relevant for practice is how good this approximation actually is.

\begin{figure*}[htb]
  \centerline{
    \includegraphics[width=0.3\textwidth,height=0.22\textwidth,angle=0]{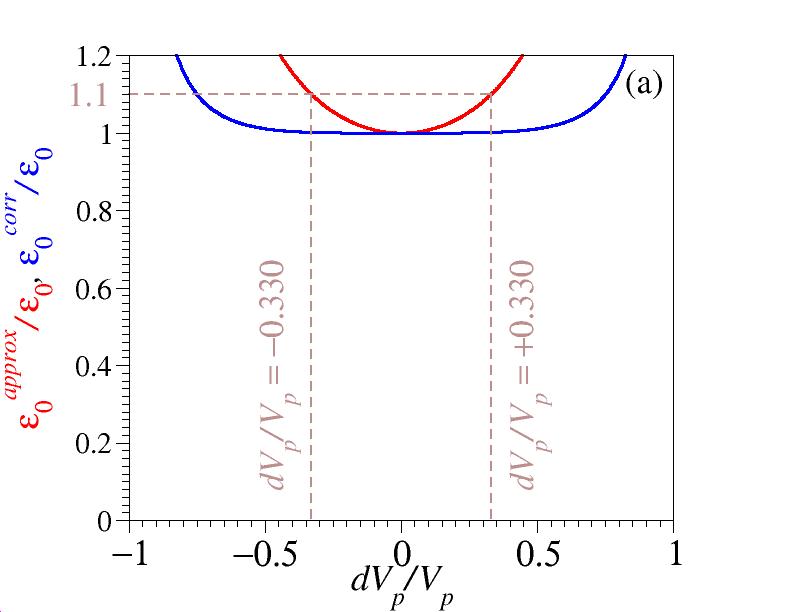}
    \includegraphics[width=0.3\textwidth,height=0.22\textwidth,angle=0]{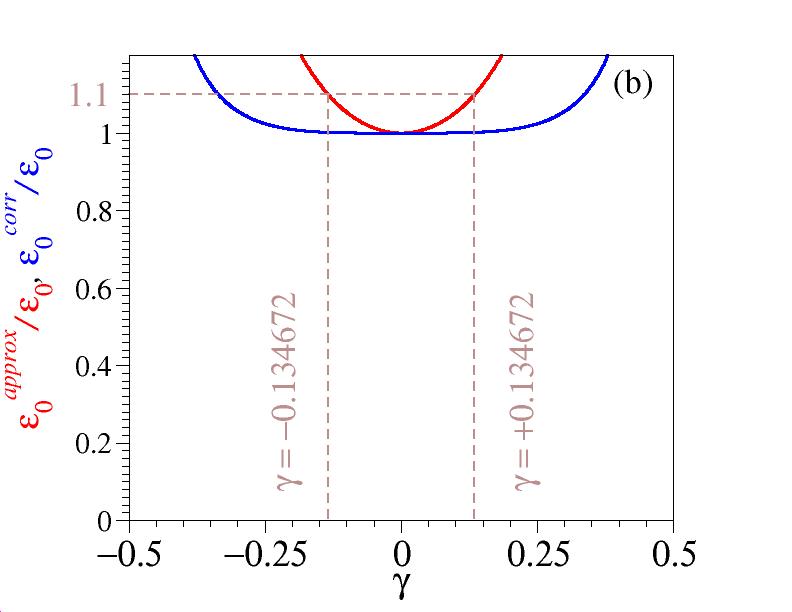}
  }
  \centerline{
    \includegraphics[width=0.3\textwidth,height=0.22\textwidth,angle=0]{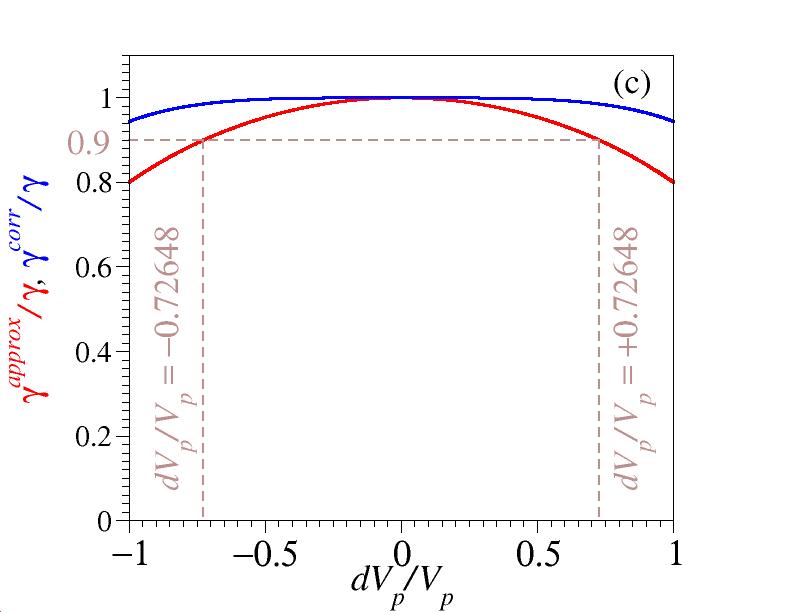}
    \includegraphics[width=0.3\textwidth,height=0.22\textwidth,angle=0]{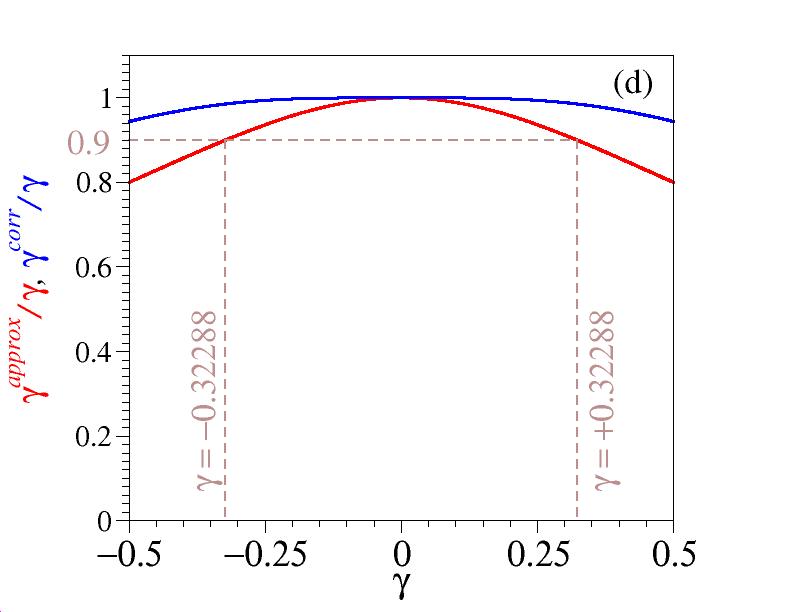}
  }
  \centerline{
    \includegraphics[width=0.3\textwidth,height=0.22\textwidth,angle=0]{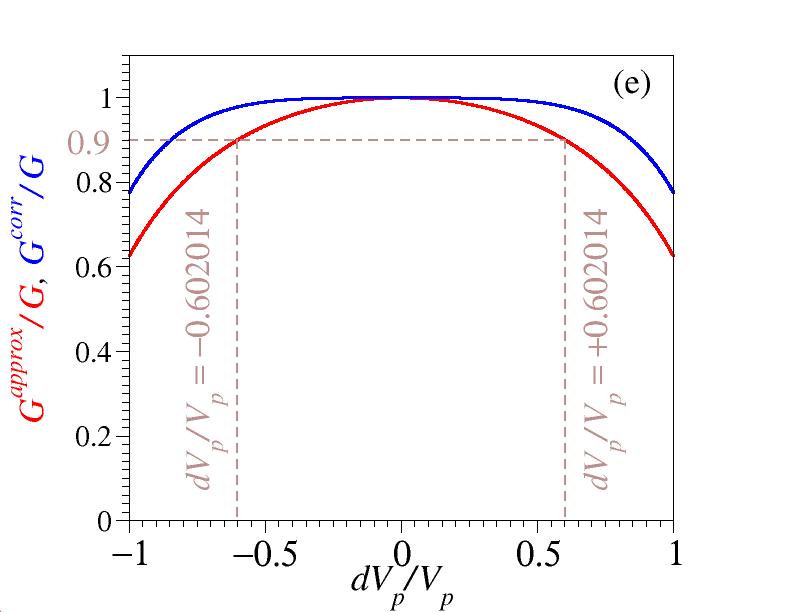}
    \includegraphics[width=0.3\textwidth,height=0.22\textwidth,angle=0]{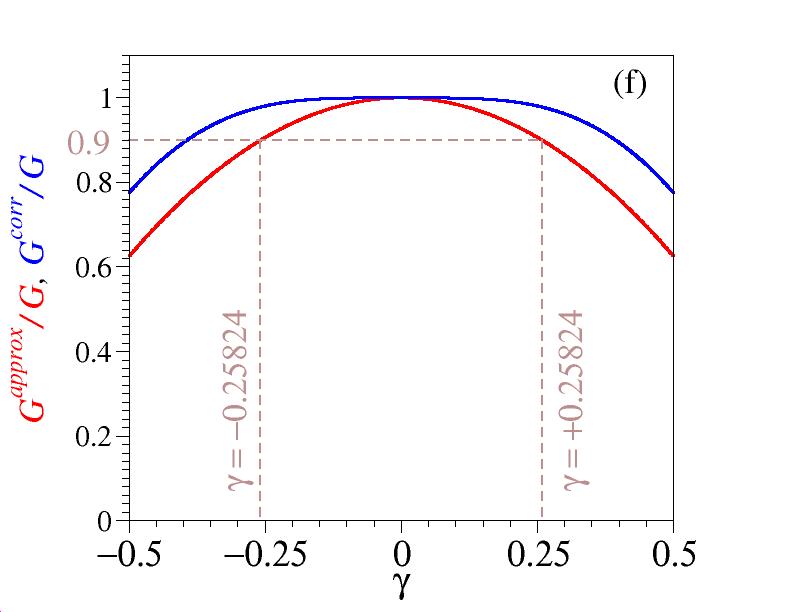}
  }
  \caption{Relative deviations from the exact values of the model parameters $\varepsilon_0$, $\gamma$, and $G$
    computed by \gl~(\ref{eq-approx}) and \gl~(\ref{eq-corr}) (red and blue lines, respectively)
    plotted versus the relative peak asymmetry location $\delta V_{p}/V_{p}$ (left panels a, c, e)
    and versus the bias-driven MO shift $\gamma$ (right panels b, d, f). The vertical brown lines visualize
    the ranges within which the parameters estimated via \gl~(\ref{eq-approx}) are accuŕate within 10\%,
    a typical value for real molecular junctions.}
  \label{fig:background}
\end{figure*}

To illustrate the accuracy of the lowest order approximation, in \figname\ref{fig:background}
I depicted by red lines departures from the exact values of the parameters  $\varepsilon_0$, $\gamma$, and $G$
estimated \gl~(\ref{eq-approx}) both as a function of the fractional peak voltage asymmetry $\delta V_{p}/V_{p}$
(panels a, c, and e) and as a function of the bias driven MO shift (panels b, d, and f).
For the reader's convenience, in \figname\ref{fig:background} I drew vertical lines
to delimit the range where the model parameters estimated
from \gl~(\ref{eq-approx}) are accurate to within 10\%, a value typical for accuracy in molecular electronics.

Inspection of \figname\ref{fig:background} reveals that all three parameters achieve this accuracy for
$\vert \gamma \vert \alt 0.134 $. 
On this basis one can conclude that \gl~(\ref{eq-approx}) is indeed a good approximation.
In saying this, I have in mind that 
for most real molecular tunnel junctions typical values of $\gamma$ are relatively small
(say $\vert \gamma\vert \alt 0.1$ \cite{Metzger:15,Metzger:16a,Baldea:2019d,Baldea:2019h,Baldea:2021d}).
This is the case even for molecular junctions exhibiting substantial current rectification.\cite{Jurchescu:23}

To better understand why the lowest order approximations of \gl~(\ref{eq-approx}) are adequate for most practical purposes
I present below the next-to-leading corrections \ib{(i.e., retaining the terms proportional to $(\delta V_p /V_p^2)$ in the Taylor series)}
\begin{subequations}
\label{eq-corr}
  \begin{eqnarray}
  \varepsilon_0 & = & \underbrace{\varepsilon_{0}^{\mbox{\small approx}}
    \left[1 - \frac{41}{50} \left(\frac{\delta V_{p}}{V_{p}}\right)^2\right]}_{\varepsilon_{0}^{\mbox{\small corr}}}
  + \mathcal{O}\left(\frac{\delta V_{p}}{V_{p}}\right)^4\\
  \gamma & = & \underbrace{\gamma^{\mbox{\small approx}} \left[1 + \frac{9}{50} \left(\frac{\delta V_{p}}{V_{p}}\right)^2\right]}_{\gamma^{\mbox{\small corr}}}
  + \mathcal{O}\left(\frac{\delta V_{p}}{V_{p}}\right)^4 \\
  G & = & \underbrace{G^{\mbox{\small approx}} \left[1 + \frac{6}{25} \left(\frac{\delta V_{p}}{V_{p}}\right)^2\right]}_{G^{\mbox{\small corr}}} 
  + \mathcal{O}\left(\frac{\delta V_{p}}{V_{p}}\right)^4
\end{eqnarray}
\end{subequations}
They are depicted by blue lines in \figname\ref{fig:background}.
As visible above, terms linear in \(\delta V_{p} / V_{p}\) vanish; only quadratic terms
\((\delta V_{p} / V_{p})^2\) contribute.

\ib{To avoid misuses, I want to explicitly emphasize what \figname\ref{fig:background} clearly visualizes.
  While the general orSLM model can be used to quantitatively analyze molecular junctions exhibiting
  strong current rectification,\cite{Metzger:18,Mondal:23}
  the five-thirds protocol cannot; it is designed to deal expeditiously with cases
  wherein current rectification is not very pronounced.}

\section{Practical recipe for the application of the five-thirds protocol}
\label{sec:recipe}

\begin{figure*}[htb]
  \centerline{
    \includegraphics[width=0.3\textwidth,height=0.22\textwidth,angle=0]{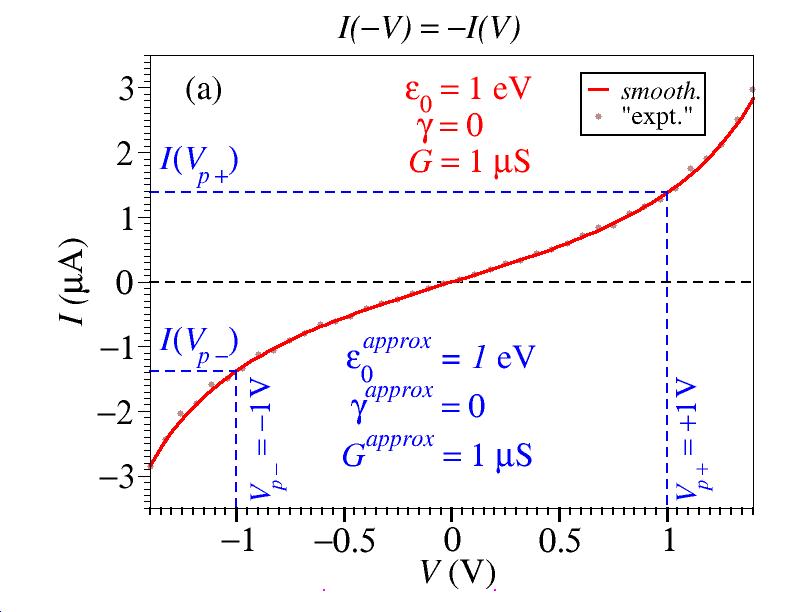}
    \includegraphics[width=0.3\textwidth,height=0.22\textwidth,angle=0]{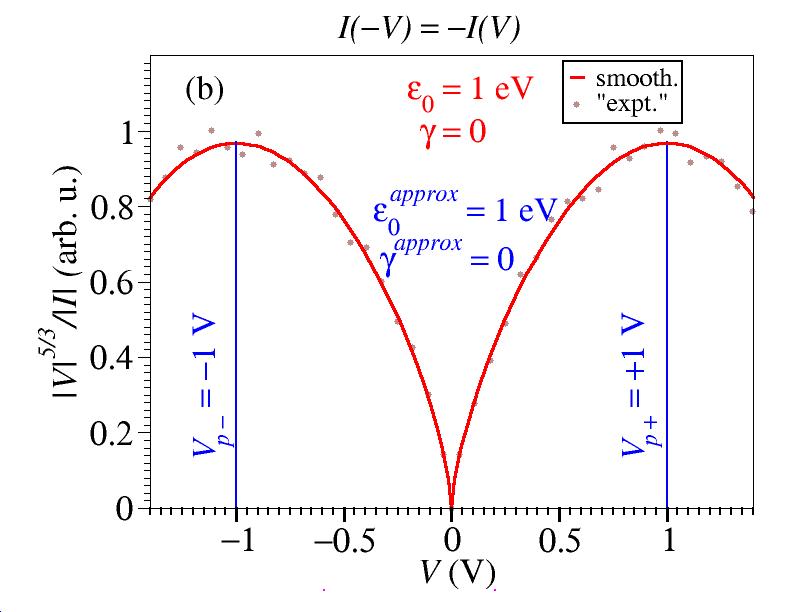}
  }
  \centerline{
    \includegraphics[width=0.3\textwidth,height=0.22\textwidth,angle=0]{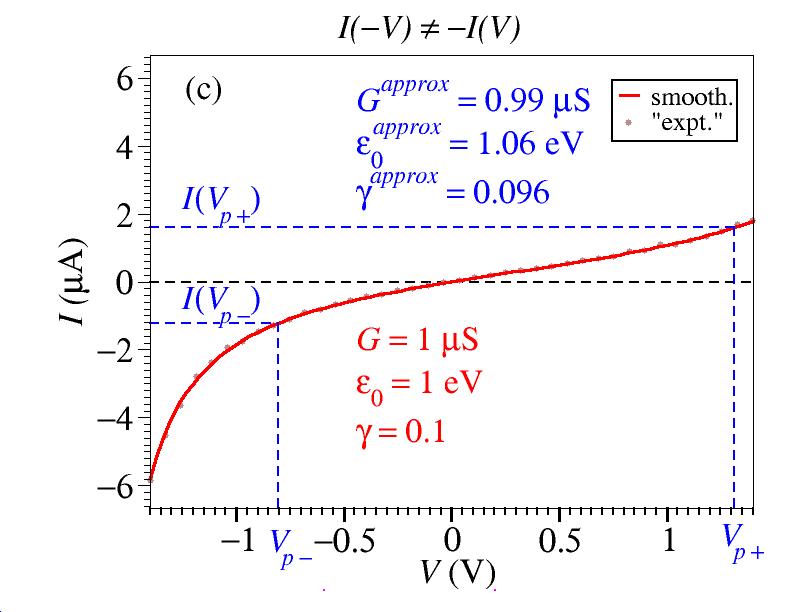}
    \includegraphics[width=0.3\textwidth,height=0.22\textwidth,angle=0]{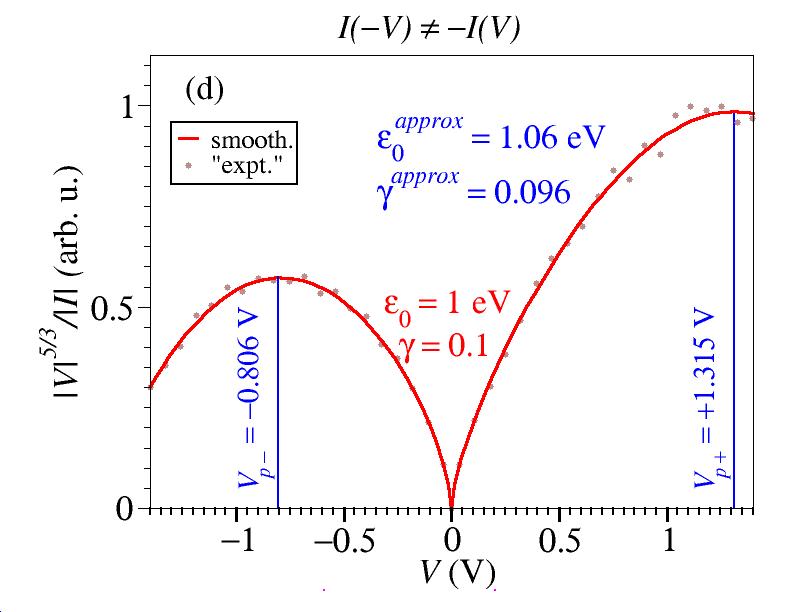}
  }
  \caption{Schematic illustration of the five-thirds protocol at work for junctions with symmetric {\iv} characteristics
    (panels a and b)
    and asymmetric {\iv} characteristics (panels c and d).
    The red curves computed by using \gl~(\ref{eq-jIB}) and parameter values depicted in red which mimic real {\iv} curves
    of panel a and c are recast as plots of $\vert V\vert^{5/3}/\vert I \vert $ versus bias ($V$) in panel b and d.
    The maximum locations at $V_{p\pm}$ extracted from the right panels
    along with the corresponding current values $I_{p\pm}$ obtained from the left panels
    are used to estimate the model parameters $\varepsilon_0$, $\gamma$, and $G$ via \gl~(\ref{eq-approx}).
    The weak disorder (brown points) overimposed on the red {\iv}
    curves is intended to give a flavor of how applications to real junctions look like.} 
  \label{fig:protocol}
\end{figure*}

\figname\ref{fig:protocol} illustrates how to apply the presently proposed five-thirds protocol to estimate the parameters for
molecular junctions possessing symmetric and asymmetric {\iv} characteristics:

(i) Recast the measured {\iv} data (panels a and c) as a plot of $\vert V\vert^{5/3}/\vert I\vert$ versus $V$ (panels b and d).

(ii) Extract the values of the peak voltages $V_{{p} +}$ and  $V_{{p} -}$ from the maxima of $\vert V\vert^{5/3}/\vert I\vert$
at positive and negative bias. These values allow the straightforward determination of the MO energy offset $\varepsilon_0$
and the MO bias-driven shift $\gamma$ via \gl~(\ref{eq-e0-approx}) and (\ref{eq-gamma-approx}), respectively.
In cases of symmetric {\iv} curves $V_{{p} + } = - V_{{p} -} = V_{{p}}$ ($\gamma = 0$) and  $\vert \varepsilon_0\vert = e V_{{p}}$. 
To accurately extract the peak positions ($V_{{p} \pm}$), noisy curves should be smoothed beforehand. This is a straightforward task
for common software utilized by experimentalists (e.g., ORIGIN).

(iii) With the values of $V_{{p} +}$ and $V_{{p} -}$ in hand, return to the {\iv} curves and extract the values of the current
$I_{{p} +}$ and $I_{{p} -}$ at the biases $V_{{p} +}$ and $V_{{p} -}$ (\figsname\ref{fig:protocol}a and c).
Use these four values ($V_{{p}\pm}$ and $I_{{p}\pm}$) to estimate
the low bias conductance from \gl~(\ref{eq-G-approx}). In cases of symmetric {\iv} curves,
$\gamma = 0$, $I_{{p} +} = -I_{{p} -} = I_{{p}}$, $V_{{p} +} = -V_{{p} -} = V_{{p}} = \vert \varepsilon_0\vert$, and \gl~(\ref{eq-G-approx})
reduces to
\begin{equation}
  \label{eq-G-5/3-symm}
  G = \frac{3}{4}\frac{I_{{p}}}{V_{{p}}}
\end{equation}

(iv) Inspection of \figname\ref{fig:background}a, c, and e allows one to assess the accuracy/reliability of the parameters
$\varepsilon_0$, $\gamma$, and $G$ estimated via \gl~(\ref{eq-approx}) at the value of $\delta V_{{p}}/V_{{p}}$
computed from the values $V_{{p} +}$ and $V_{{p} -}$ directly extracted from the experimental {\iv} data in question
without any assumption.

\section{Applications to real junctions}
\label{sec:applications}

To emphasize the generality of the presently proposed five-thirds protocol, I will consider junctions fabricated
with the most frequently utilized fabrication platforms: single-molecule mechanically-controlled break junctions,
single-molecule STM break junctions ($N = 1$), CP-AFM molecular junctions ($N \sim 100$), and large area
molecular junctions ($N \gg 1$).

As noted on several occasions (e.g., ref.~\citenum{Baldea:2024a}) {\iv} data alone do not suffice to specify
whether conduction is of p-type ($\varepsilon_0 < 0$) or n-type ($\varepsilon_0 > 0$)
(i.e., mediated by an occupied or an unoccupied MO, respectively). However,
in order to simplify the analysis of the real junctions considered below, I will ``postulate''
throughout p-type conduction. If the contrary holds true, the values of $\varepsilon_0$ and $\gamma$ determined below
should be replaced by $-\varepsilon_0$ and $-\gamma$.

\subsection{Mechanically controlled break junctions}
\label{sec:mc-bj}
\begin{figure*}[htb]
  \centerline{
    \includegraphics[width=0.3\textwidth,height=0.22\textwidth,angle=0]{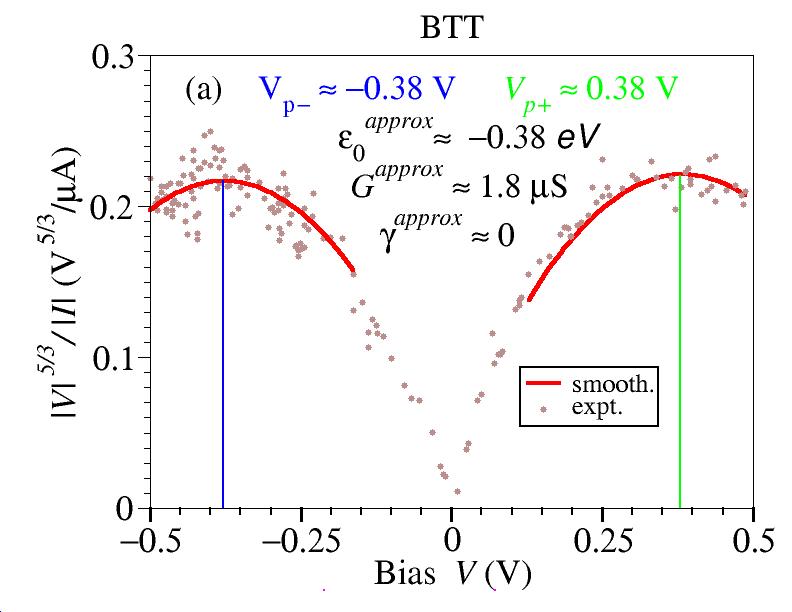}
    \includegraphics[width=0.3\textwidth,height=0.22\textwidth,angle=0]{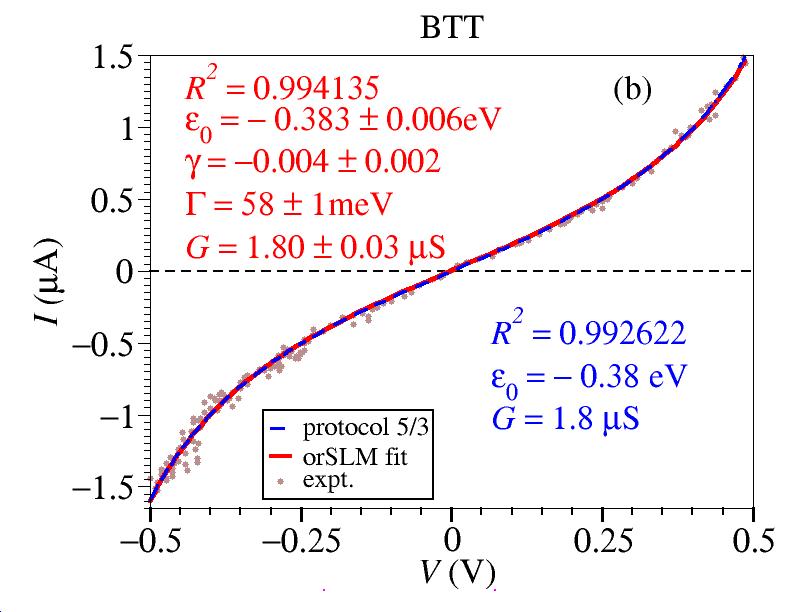}
  }
    \centerline{
    \includegraphics[width=0.3\textwidth,height=0.22\textwidth,angle=0]{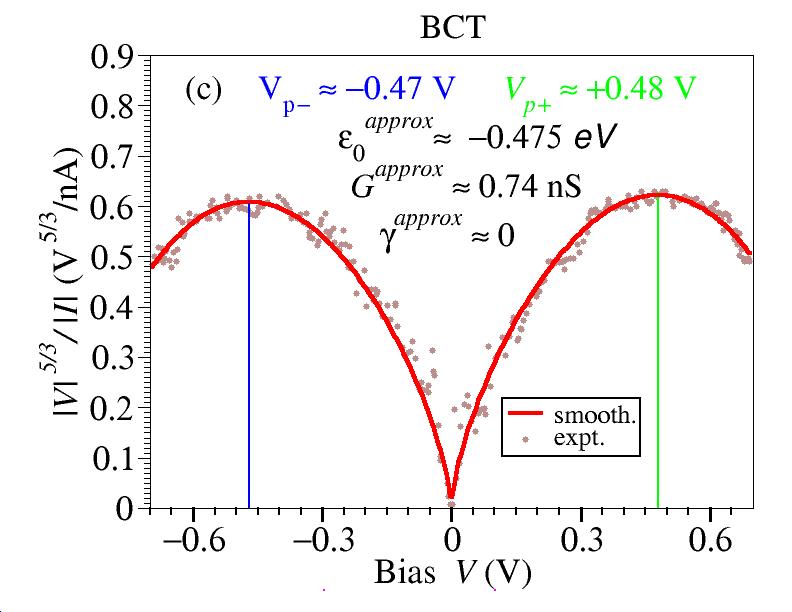}
    \includegraphics[width=0.3\textwidth,height=0.22\textwidth,angle=0]{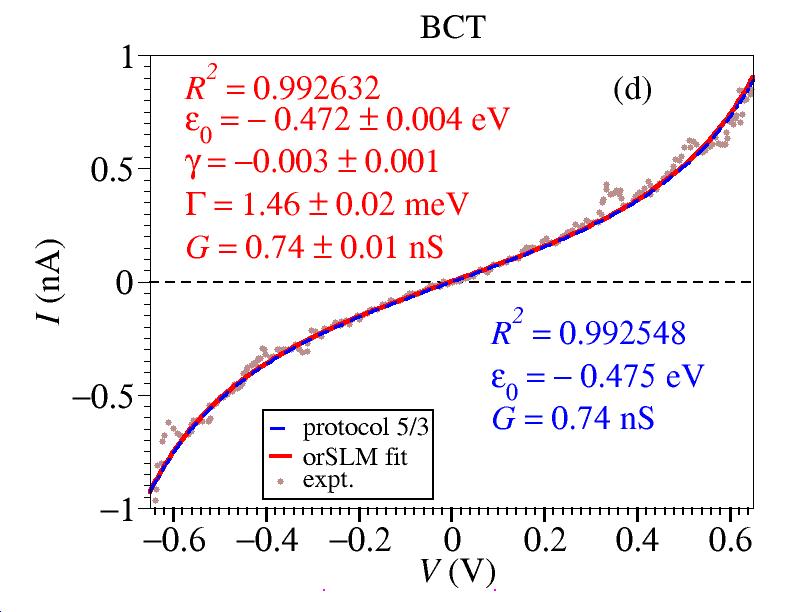}
  }
    \caption{Application of the five-thirds protocol to mechanically controlled break junctions fabricated with
      (4,4'-bisthiotolane (BTT)) (panels a and b) and 4,4'-biscyanotolane (BCT) (panels c and d).
      The experimental {\iv} data (brown points) were obtained by digitizing experimental {\iv} curves reported in ref.~\citenum{Zotti:10}.
      Smoothing of the experimental data (brown points) yielded the (red) curves in panels a and c, which allow
      the reliable extraction of the biases $V_{p+}$ and $V_{p-}$ at the peaks of the $\vert V\vert^{5/3}/\vert I\vert$ curves.
      The values $V_{p+}$ and $V_{p-}$ and the corresponding currents
      $I_{p+} \equiv I(V_{p+})$ and $I_{p-}  \equiv I(V_{p-})$ deduced from
      the {\iv} curves (panel b and d) serve as input in \gl~(\ref{eq-approx}).
      The results thus obtained ({\iv} curves and parameter values depicted in blue in panels b and d)
      have a quality comparable to those (depicted in red)
      deduced via data fitting to \gl~(\ref{eq-jIB}) with adjustable parameters. The values of $R^2$ depicted in blue
      were computed via \gl~(\ref{eq-r2}).
    }
  \label{fig:mc-bj}
\end{figure*}

To start with, I will consider two single-molecule mechanically controlled break junctions fabricated with tolane
anchored on gold electrodes using thiol and cyano groups (4,4'-bisthiotolane (BTT)) and 4,4'-biscyanotolane (BCT),
respectively.\cite{Zotti:10}
Digitized {\iv} curves for these junctions from ref.~\citenum{Zotti:10} are depicted by brown points in \figname\ref{fig:mc-bj}b and d.

Due to the substantial noise of the experimental (digitized) curves (brown points in \figname\ref{fig:mc-bj}a and c),
data smoothing
(red curves in \figname\ref{fig:mc-bj}a and c) represents the first
step needed in reliably extracting the bias values $V_{p +}$ and $V_{p -}$ at the peaks of the curve for
$\vert V\vert^{5/3}/\vert I\vert $.

Given the fact that the {\iv} curve for the symmetric BTT molecule is practically symmetric, the maxima are located symmetric
around origin $V_{p + } \approx - V_{p -} \approx 0.38$\,V.
With the value $I_p = \vert I(V_{p\pm}\vert \approx 0.68$\,nA) estimated from
the experimental {\iv} curve (\figname\ref{fig:mc-bj}a) the conductance at low bias $ G \approx 1.8$\,nS can be deduced using
\gl~(\ref{eq-G-5/3-symm}).

The {\iv} curve (brown points in \figname\ref{fig:mc-bj}d)
for the asymmetric BCT molecule is slightly asymmetric. Accordingly, the peaks of the curve for
$\vert V\vert^{5/3}/\vert I\vert $ are located slightly asymmetric around origin
($V_{p + } \approx 0.48$\,V and $V_{p -} \approx -0.47$\,V, cf.~\figname\ref{fig:mc-bj}c).
With these vales, the parameters $\varepsilon_0$, $\gamma$, and $G$ of the BCT junction
shown in the inset of \figname\ref{fig:mc-bj}c were estimated using \gl~(\ref{eq-approx}).

To illustrate the reliability of the five-thirds protocol for the BTT and BCT junctions considered,
along the experimental {\iv} curve (brown points), I present both
the fitting curve (line and parameter values pertaining to it in \figname\ref{fig:mc-bj}b and d are depicted in red)
obtained using \gl~(\ref{eq-jIB}) with adjustable model parameters
and the {\iv} curve (line  and parameter values pertaining to it in \figname\ref{fig:mc-bj}b and d are depicted in blue)
computed via \gl~(\ref{eq-jIB}) with the model parameters $\varepsilon_0$, $\gamma$, and $G$
provided by the five-thirds protocol. As visible, 
the (blue) curves based on the five-thirds protocol and the fitting (red) curves 
cannot be distinguished from each other within the drawing accuracy.
To put this excellent agreement in more quantitative terms,
along with the coefficient of determination $R^2$ obtained by data fitting to 
\gl~(\ref{eq-jIB}) (values written in red in \figname\ref{fig:mc-bj}b and d),
I also present the counterparts of $R^2$ for the five-thirds protocol
(values written in blue in \figname\ref{fig:mc-bj}b and d).
The latter was obtained in the standard manner
\begin{subequations}
  \label{eq-r2}
\begin{eqnarray}
  R^2 & = & 1 - SS_{\mbox{\small res}} / SS_{\mbox{\small tot}} \\
  SS_{\mbox{\small res}} & = & \sum_{k=1}^{n}\left[I_{k}^{\mbox{\small exp}} - I\left(V_k; \varepsilon_0^{approx}, \gamma^{approx}, G^{approx}\right)\right]^2 \\
  SS_{\mbox{\small tot}} & = & \sum_{k=1}^{n}\left(I_{k}^{\mbox{\small exp}} - \overline{I}\right)^2; \overline{I} = \frac{1}{n} \sum_{k_=1}^{n} I_{k}^{\mbox{\small exp}}
\end{eqnarray}
\end{subequations}
with $n$ experimental values of the current $I_k^{\mbox{\small exp}}$ and the values
$I\left(V_k; \varepsilon_0^{approx}, \gamma^{approx}, G^{approx}\right)$ computed
from \gl~(\ref{eq-jIB}) at the biases $V_k$ sampled in experiment
using the values $ \varepsilon_0^{approx}, \gamma^{approx}, G^{approx} $ of \gl~(\ref{eq-approx}).

\subsection{STM break junctions}
\label{sec:stm}
\begin{figure*}[htb]
  \centerline{
    \includegraphics[width=0.3\textwidth,height=0.22\textwidth,angle=0]{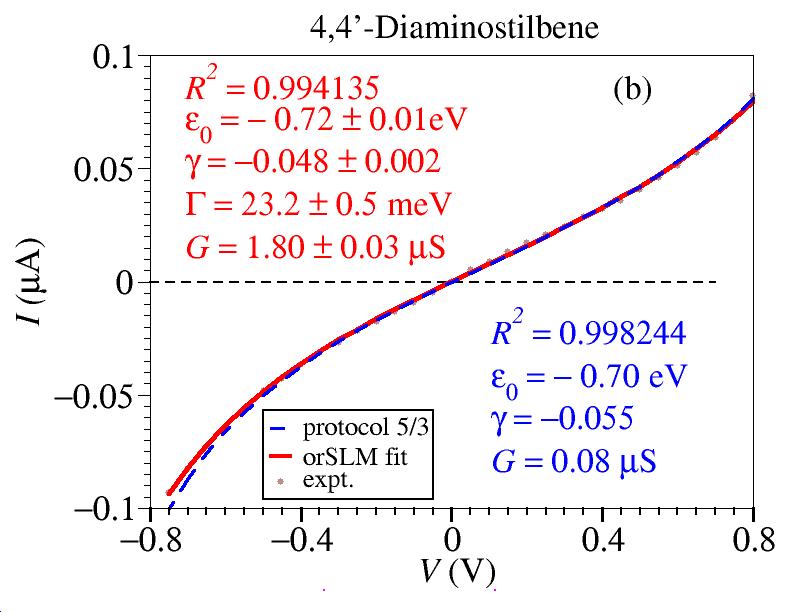}
    \includegraphics[width=0.3\textwidth,height=0.22\textwidth,angle=0]{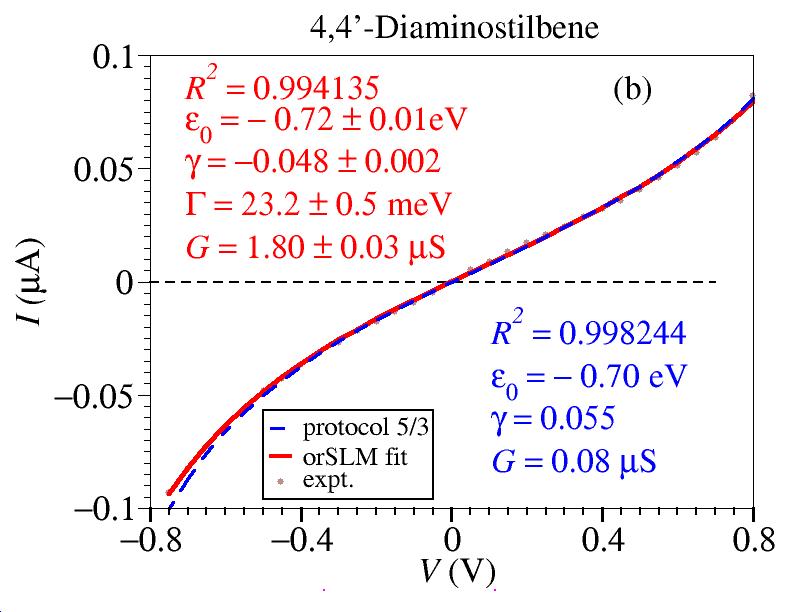}
    \includegraphics[width=0.3\textwidth,height=0.22\textwidth,angle=0]{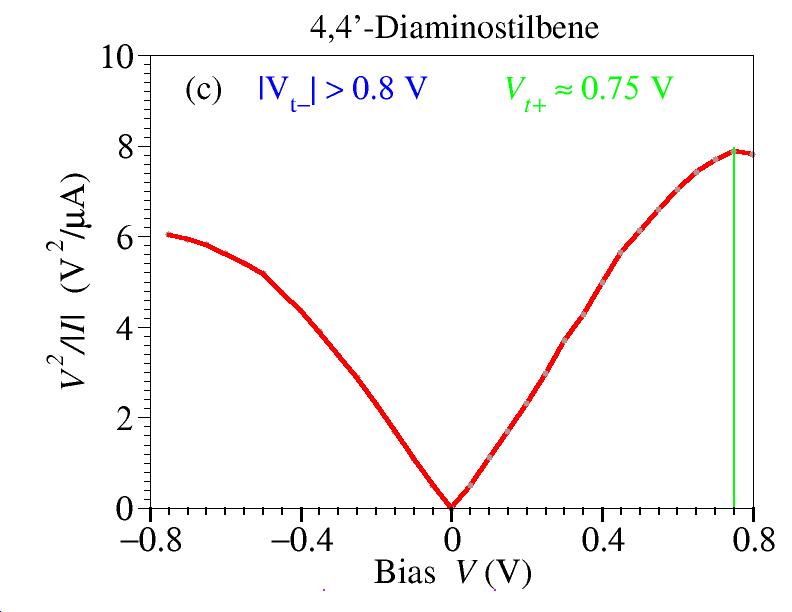}
  }
   \centerline{
    \includegraphics[width=0.3\textwidth,height=0.22\textwidth,angle=0]{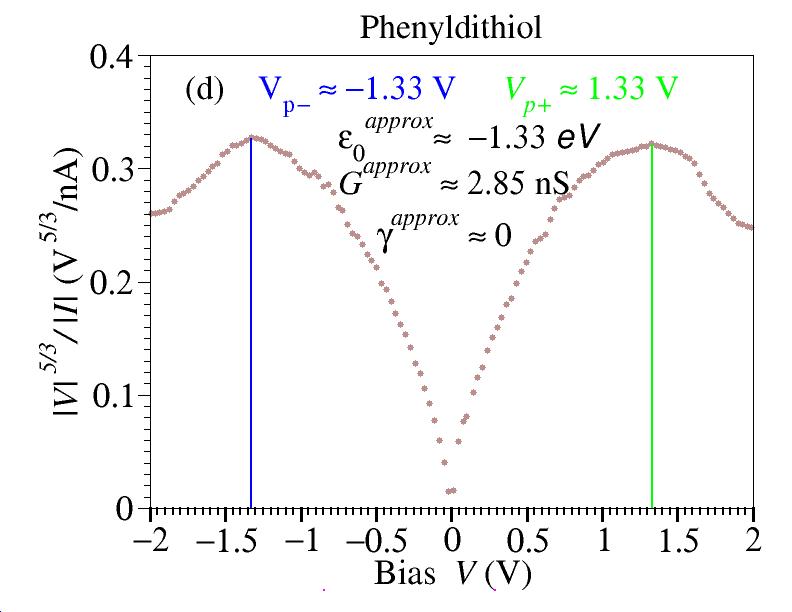}
    \includegraphics[width=0.3\textwidth,height=0.22\textwidth,angle=0]{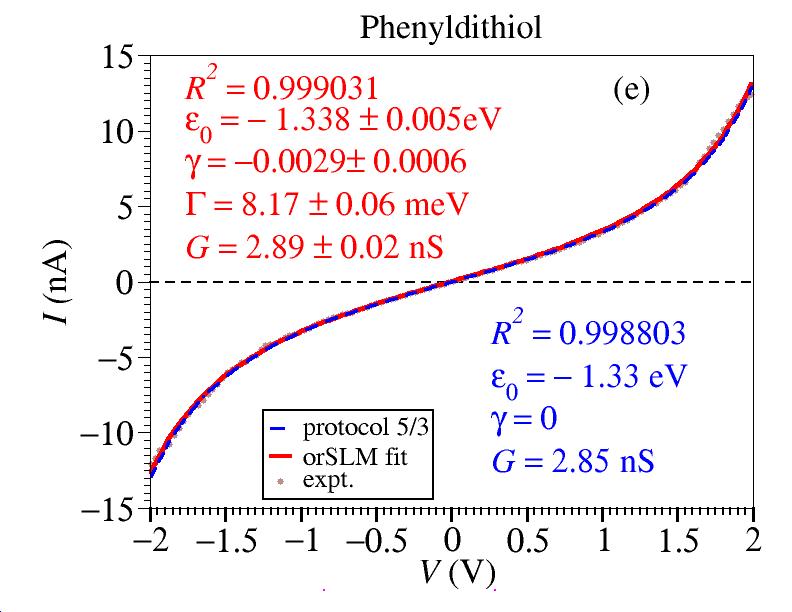}
  } 
   \caption{Application of the five-thirds protocol to STM break junctions fabricated with
     4,4'-diaminostilbene \cite{Venkataraman:09a} (panels a, b, and c) and phenyldithiol (panel d and e).\cite{Reddy:11}
     The values $V_{p+}$ and $V_{p-}$ at the peaks of the $\vert V\vert^ {5/3}/\vert I\vert$ curves
     and the corresponding currents
      $I_{p+} \equiv I(V_{p+})$ and $I_{p-}  \equiv I(V_{p-})$ deduced from
      the {\iv} curves (panel b and e) serve as input in \gl~(\ref{eq-approx}).
      The results thus obtained ({\iv} curves and parameter values depicted in blue in panels b and e)
      have a quality comparable to those (depicted in red)
      deduced via data fitting to \gl~(\ref{eq-jIB}) with adjustable parameters. The values of $R^2$ depicted in blue
      were computed via \gl~(\ref{eq-r2}). Panel c illustrates that the present five-thirds protocol can be also be applied in situations
      where the bias range sampled in experiment is too narrow for applying TVS.\cite{Beebe:06,Baldea:2019d,Baldea:2019h}}
  \label{fig:stm-bj}
\end{figure*}

Next I will examine two single-molecule STM-BJ junctions with gold electrodes fabricated using 
4,4'-diaminostilbene \cite{Venkataraman:09a} (\figname\ref{fig:stm-bj}a to c) and phenyldithiol \cite{Reddy:11} (\figname\ref{fig:stm-bj}d and e).

In both cases no data smoothing was required to reliably extract $V_{p +}$ and $V_{p-}$ from the peaks of \figsname\ref{fig:stm-bj}a and d.
With these values and the pertaining currents $I_{p +}$ and $I_{p-}$ deduced from the experimental {\iv} curves of \figsname\ref{fig:stm-bj}b and e
(brown points), I arrived via \gl~(\ref{eq-approx}) at the blue {\iv} curves. Again, these curves excellently agree with the
red {\iv} curves obtained by data fitting to \gl~(\ref{eq-jIB}) with model parameters adequately adjusted.

The junction of 4,4'-diaminostilbene allows one to reveal the potential advantage of the present five-thirds protocol over
the standard TVS-orSLM approach \cite{Baldea:2012a,Baldea:2019d,Baldea:2019h} based on the transition voltage $V_t$.
As visible in \figsname\ref{fig:stm-bj}c, the range of negative biases sampled in experiment \cite{Venkataraman:09a} was not sufficiently
broad. This prevents the determination of the model parameters using \gl~(\ref{eq-tvs}). That is, the five-thirds protocol
can also be applied in cases where the use of \gl~(\ref{eq-tvs}) is impractical.

\subsection{CP-AFM molecular junctions}
\label{sec:cp-afm}
\begin{figure*}[htb]
   \centerline{
    \includegraphics[width=0.3\textwidth,height=0.22\textwidth,angle=0]{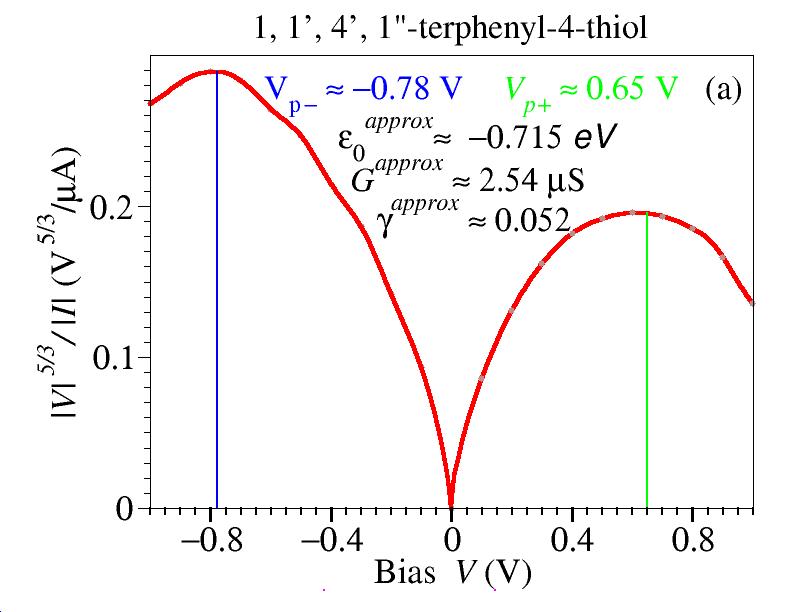}
    \includegraphics[width=0.3\textwidth,height=0.22\textwidth,angle=0]{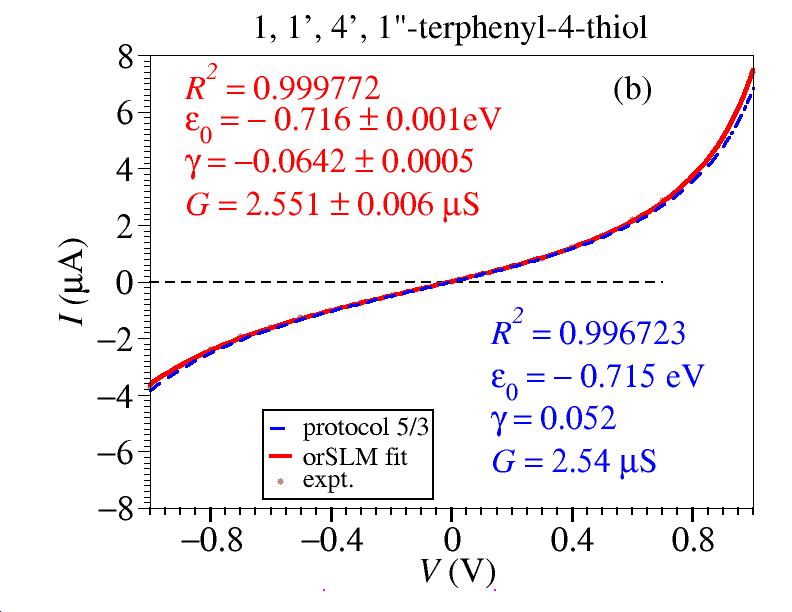}
   }
   \centerline{
     \includegraphics[width=0.3\textwidth,height=0.22\textwidth,angle=0]{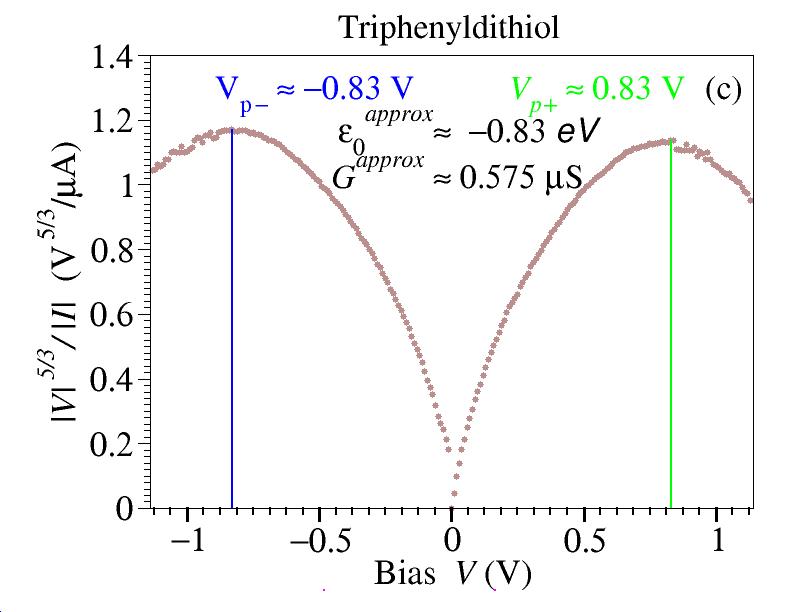}
     \includegraphics[width=0.3\textwidth,height=0.22\textwidth,angle=0]{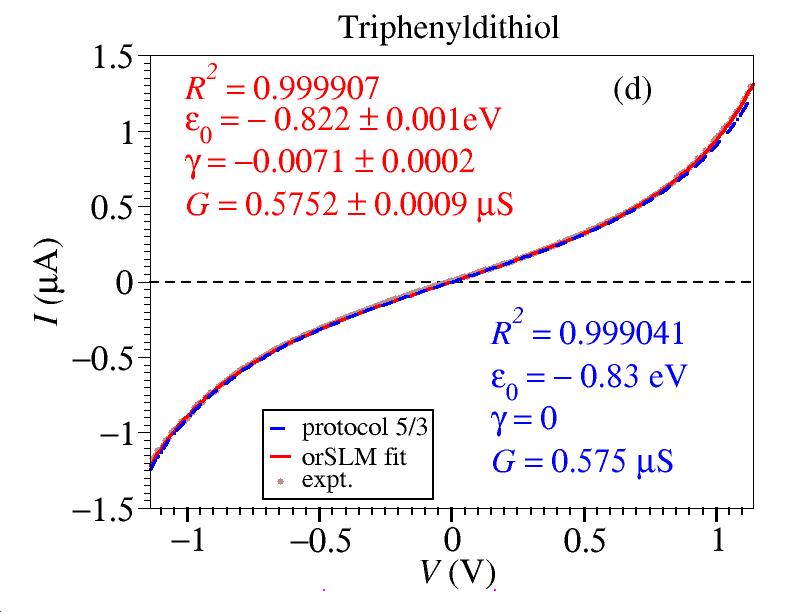}
   }
   \centerline{
     \includegraphics[width=0.3\textwidth,height=0.22\textwidth,angle=0]{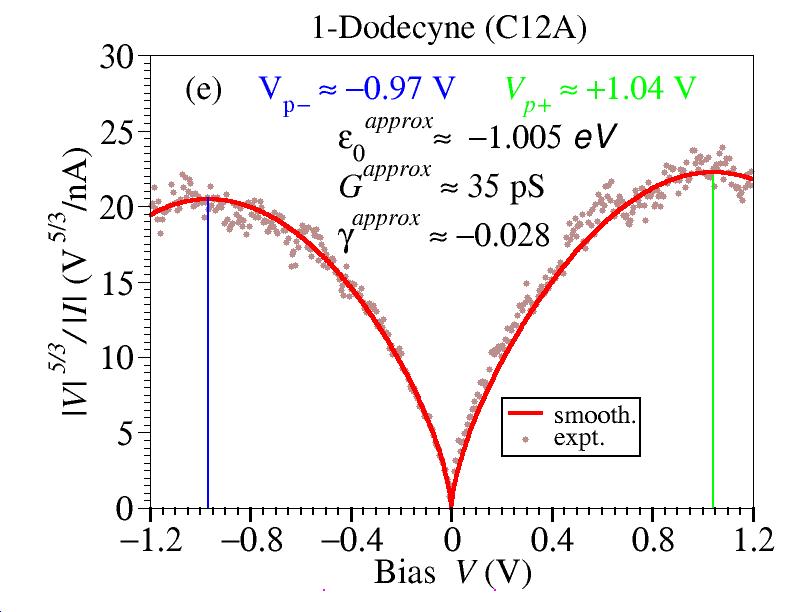}
     \includegraphics[width=0.3\textwidth,height=0.22\textwidth,angle=0]{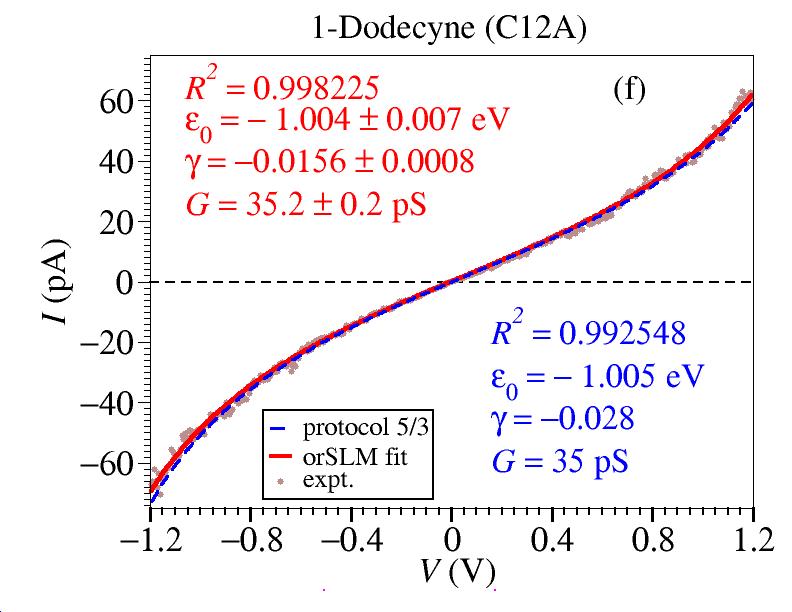}
   }    
   \caption{Application of the five-thirds protocol to CP-AFM junctions fabricated with
     1, 1'-, 4', 1''-terphenyl-4-thiol and gold electrodes (panels a and b),\cite{Tan:10}
     with triphenyldithiol and silver electrodes (panels c and d),\cite{Baldea:2015d}
     \ib{and with 1-dodecyne and silver electrodes (panel e and f).\cite{Baldea:2024d}}
     The values $V_{p+}$ and $V_{p-}$ at the peaks of the $\vert V\vert^ {5/3}/\vert I\vert$ curves
     and the corresponding currents
      $I_{p+} \equiv I(V_{p+})$ and $I_{p-}  \equiv I(V_{p-})$ deduced from
      the {\iv} curves (panel b and e) serve as input in \gl~(\ref{eq-approx}).
      The results thus obtained ({\iv} curves and parameter values depicted in blue in panels b and e)
      have a quality comparable to those (depicted in red)
      deduced via data fitting to \gl~(\ref{eq-jIB}) with adjustable parameters. The values of $R^2$ depicted in blue
      were computed via \gl~(\ref{eq-r2}).
      The experimental data of panels d (junctions based on triphenyldithiols)
      and \ib{f (junctions based on 1-dodecyne (C12A))}
      were measured in conjunction with work reported in
      refs.~\citenum{Baldea:2015d} and \ib{\citenum{Baldea:2024d} (courtesy of Zuoti Xie)}.}
  \label{fig:cp-afm}
\end{figure*}

Smoothing the experimental transport data
is also superfluous in analyzing the CP-AFM junctions of 1,1'-,4',1''-terphenyl-4-thiol 
and gold electrodes \cite{Tan:10} (\figsname\ref{fig:cp-afm}a and b)
and of triphenyldithiol and silver electrodes (\figsname\ref{fig:cp-afm}c and d).\cite{Baldea:2015d}
\ib{Data smoothing is necessary to process the experimental {\iv} curve 
  measured for the recently investigated CP-AFM junctions fabricated with 1-dodecyne
  (C12A) and silver electrodes anchored via alkynyl groups depicted in \figsname\ref{fig:cp-afm}e and f.\cite{Baldea:2024d}}

Inspection of the parameter values in the legends reveals that the five-thirds protocol
works for all these cases.  

\subsection{Large-area molecular junctions}
\label{sec:large-area}

\begin{figure*}[htb]
  \centerline{
    \includegraphics[width=0.3\textwidth,height=0.22\textwidth,angle=0]{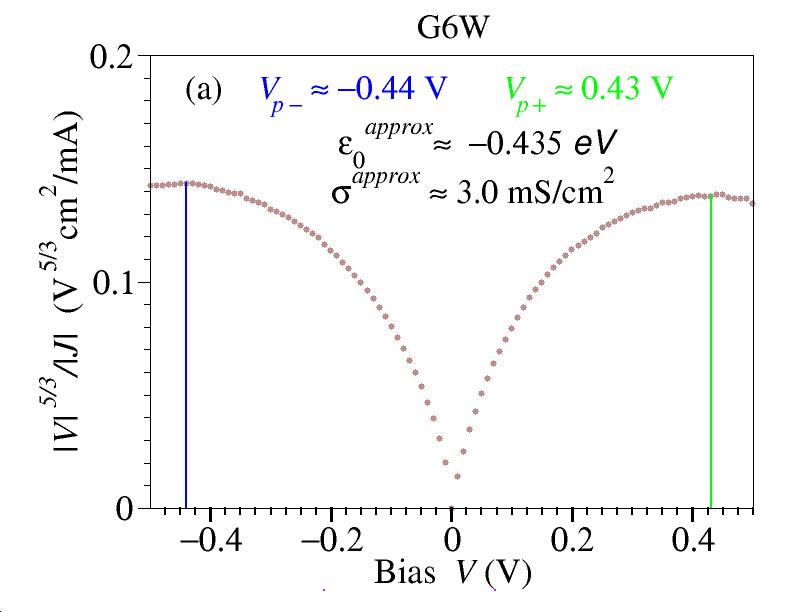}
    \includegraphics[width=0.3\textwidth,height=0.22\textwidth,angle=0]{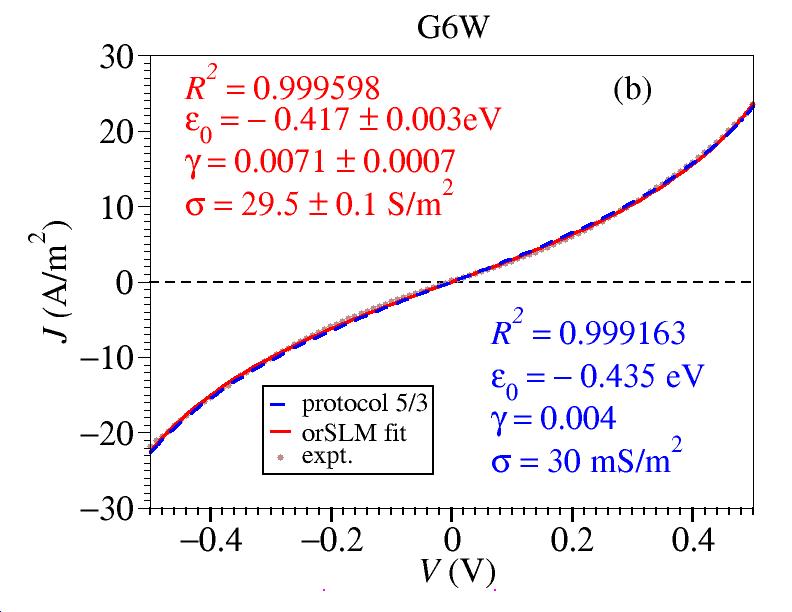}
    \includegraphics[width=0.3\textwidth,height=0.22\textwidth,angle=0]{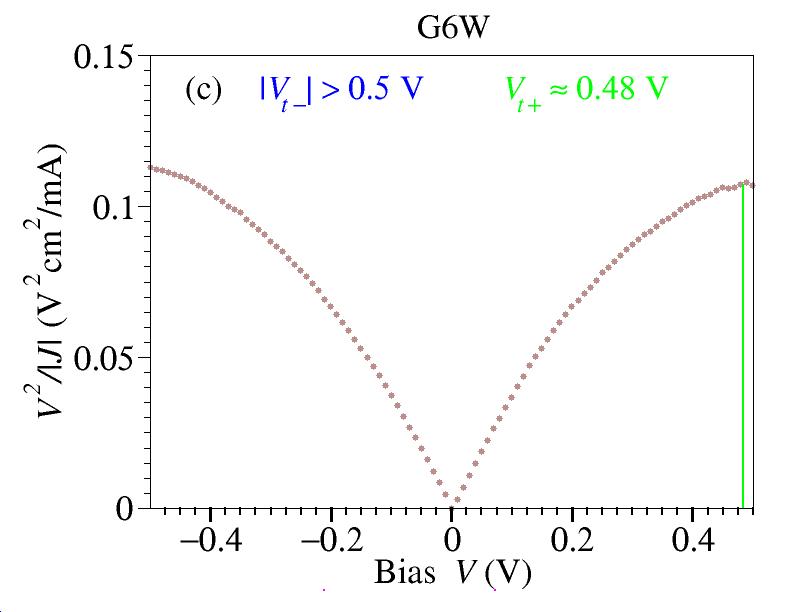}
  }
   \centerline{
     \includegraphics[width=0.3\textwidth,height=0.22\textwidth,angle=0]{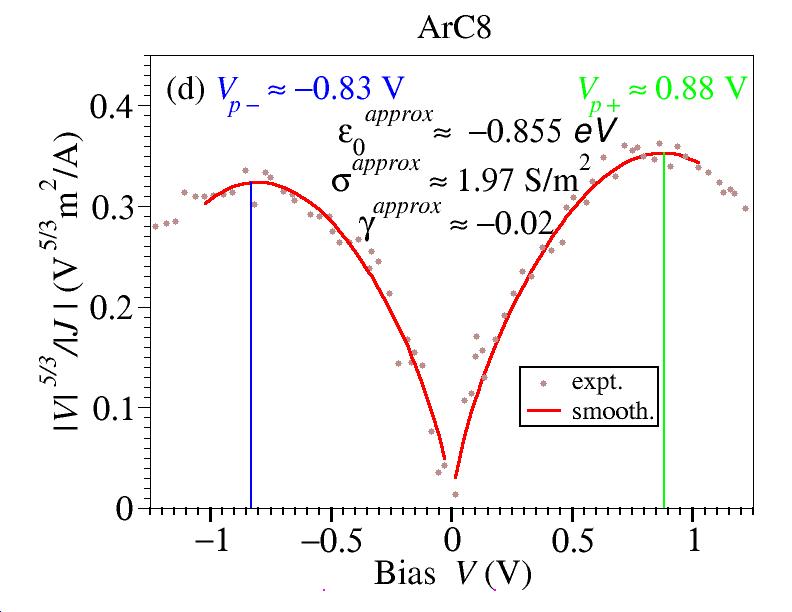}
     \includegraphics[width=0.3\textwidth,height=0.22\textwidth,angle=0]{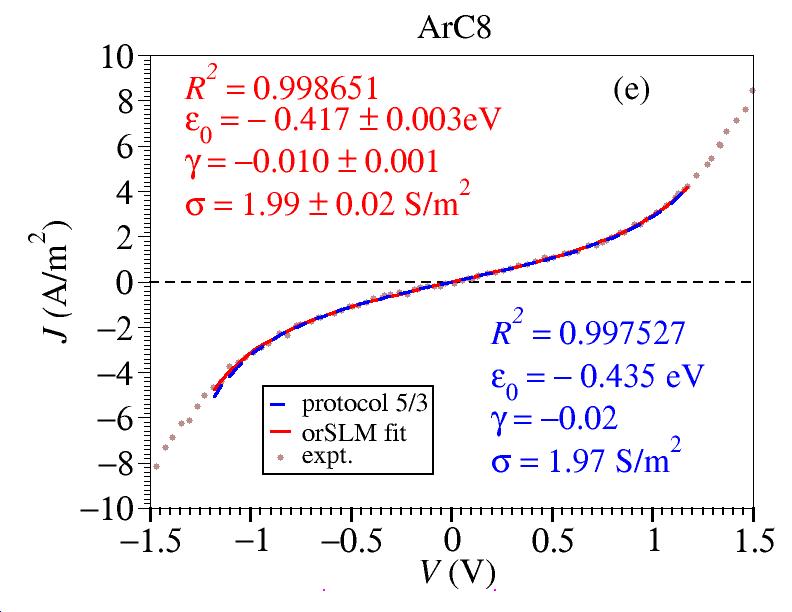}
   }
   \centerline{
    \includegraphics[width=0.3\textwidth,height=0.22\textwidth,angle=0]{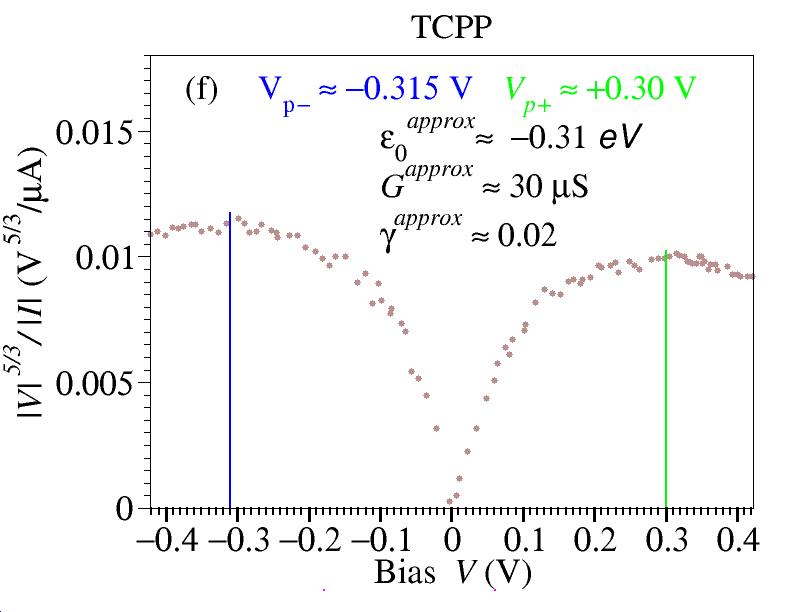}
    \includegraphics[width=0.3\textwidth,height=0.22\textwidth,angle=0]{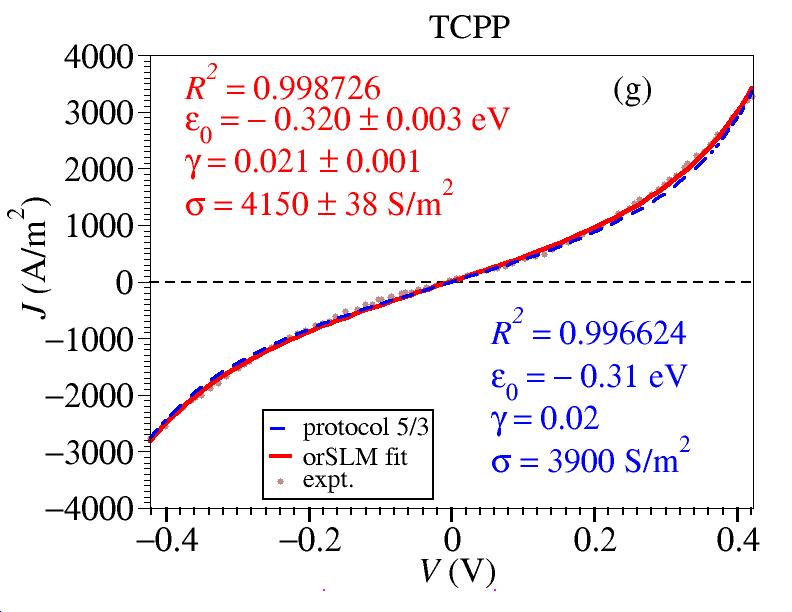}
   }
   \caption{Application of the five-thirds protocol to large area molecular junctions fabricated with
     a peptide (G6W) \cite{Su:23} (panels a, b, and c; experimental data: courtesy of Cunlan Guo),
     aryl octane (ArC8) \cite{Song:22b} (panel d and e; experimental data from digitized \figname{4a} of ref.~\citenum{Song:22b}),
     \ib{and metal free ITO-based TCPP (panels f and g; experimental from digitized \figname{4c} of ref.~\citenum{Sergani:13})}.
     The values $V_{p+}$ and $V_{p-}$ at the peaks of the $\vert V\vert^ {5/3}/\vert I\vert$ curves
     and the corresponding currents
      $I_{p+} \equiv I(V_{p+})$ and $I_{p-}  \equiv I(V_{p-})$ deduced from
      the {\iv} curves (panel b, e \ib{and g}) serve as input in \gl~(\ref{eq-approx}).
      The results thus obtained ({\iv} curves and parameter values depicted in blue in panels b, e \ib{and g})
      have a quality comparable to those (depicted in red)
      deduced via data fitting to \gl~(\ref{eq-jIB}) with adjustable parameters. The values of $R^2$ depicted in blue
      were computed from \gl~(\ref{eq-r2}).
      Panel c illustrates that the present five-thirds protocol can be also be applied in situations
      where the bias range sampled in experiment is too narrow for applying TVS.\cite{Beebe:06,Baldea:2019d,Baldea:2019h}
      Because the transport measurements for large area junctions report current densities ($J$) rather than current intensities ($I$),
      values of the conductivity $\sigma$ rather that conductance $G$ are presented in this figure.
      In analyzing the ArC8 \cite{Song:22b} \ib{and TCPP \cite{Sergani:13} data}, the bias range
      considered has been restricted to the bias range for which the
      orSLM applies.\cite{Baldea:2012a,Baldea:2023a,Baldea:2024a}}
  \label{fig:large-area}
\end{figure*}

Last but not least, I will focus on \ib{three} large area molecular junctions \ib{completely different
from each other} (\figname\ref{fig:large-area}).

The results depicted in \figsname\ref{fig:large-area}a to c refer to a 
peptide-based junction fabricated with gold substrate and EGaIn top electrodes.\cite{Su:23}
The specific peptide considered (G6W = GGGGGGW) consists of six glycines (G) with one aromatic amino acid at
the C-terminus (tryptophan, W).
\figsname\ref{fig:large-area}d and e pertain to a junction consisting of a self-assembled monolayer of
aryl octane (ArC8) with graphene contacts as protecting interlayer.
\ib{The results depicted in \figsname\ref{fig:large-area}f and g are for 
  metal-free ITO-TCPP/PEDOT:PSS molecular junctions.\cite{Sergani:13}
  They were fabricated using carboxylic acid-modified porphyrin (meso-tetra(4-carboxyphenyl)porphyrin, TCPP) adsorbed to a 
  bottom electrode of indium tin oxide (ITO) and having the conductive PEDOT:PSS
  (poly(3,4-ethylenedioxythiophene):poly(styrenesulfonate))
  polymer as top electrode.\cite{Sergani:13}
  Recall that ITO is a degenerate n-type semiconductor possessing a wide band gap which makes it a
  transparent conductive electrode routinely employed in optoelectronic devices.}

The message conveyed by the numerical values of the parameters inserted in \figname\ref{fig:large-area}
should be obvious. As in the preceding cases, they validate the five-thirds protocol
also for the large-area junctions considered. To avoid some misunderstandings persisting in the literature,
validation of the five-thirds protocol implicitly validates the orSLM for large area molecular tunnel junctions,
the model on which this protocol relies. In addition, \figname\ref{fig:large-area}c
reveals the same advantage of the the five-thirds protocol over the conventional TVS approach based
on \gl~(\ref{eq-tvs}) already noted in the discussion related to \figname\ref{fig:stm-bj}c:
to be applicable, the five-thirds protocol requires a narrower bias range than needed for TVS.

\section{Additional remarks}
\label{sec:remarks}

For completeness, let me finally mention that, similar to \gl~(\ref{eq-approx}),
lowest expansions in the location asymmetry ($ \delta V_{\kappa} $) of the peaks 
$V_{\kappa + } = V_{\kappa} + \delta V_{\kappa}$ and $V_{\kappa - } = - V_{\kappa} + \delta V_{\kappa}$ 
of the general quantity $\vert V\vert^{\kappa}/\vert I\vert $ can also be deduced
\begin{subequations}
\label{eq-asymm-kappa}
  \begin{eqnarray}
  \vert \varepsilon_0 \vert & \approx & \sqrt{\frac{\kappa + 1}{\kappa - 1}} \, \frac{e V_{{\kappa}}}{2} \\
  \gamma & \approx & \frac{\sqrt{\kappa^2 - 1}}{2 \kappa} \frac{\delta V_{{\kappa}}}{V_{{\kappa}}} \mbox{sign}\, \varepsilon_0\\
  G & \approx & \frac{1}{\kappa + 1}
  \left( \frac{I_{{ \kappa } +}}{V_{{ \kappa } +}} + \frac{I_{{ \kappa } -}}{V_{{ \kappa } +}} \right)
\end{eqnarray}
\end{subequations}

In the particular case $\kappa = 2$ (``transition voltage spectroscopy'', TVS),
they read
\begin{subequations}
\label{eq-asymm-2}
  \begin{eqnarray}
    V_{t\pm} & \equiv & \pm V_t + \delta V_t,\
    V_t = \frac{V_{t+} - V_{t-}}{2},\
    \delta V_t = \frac{V_{t+} + V_{t-}}{2} \nonumber \\
 \varepsilon_0 & \approx & \varepsilon_0^{\mbox{\small approx}} = \frac{\sqrt{3}}{2}e  V_t =
  e \frac{\sqrt{3}}{2}\frac{V_{{ t } +} + \vert V_{{ t } -}\vert }{2}\\
  \gamma & \approx & \gamma^{\mbox{\small approx}} = \frac{\sqrt{3}}{4} \frac{\delta V_{t}}{V_{t}} \mbox{sign}\, \varepsilon_0 \nonumber \\
  & = & \frac{\sqrt{3}}{2} \frac{V_{{ t } +} - \vert V_{{ t } -}\vert}{V_{{ t } +} + \vert V_{{ t } -}\vert} \mbox{sign}\, \varepsilon_0 \\
  G & \approx & G^{\mbox{\small approx}} = \frac{1}{3}\left( \frac{I_{{ t } +}}{V_{{ t } +}} + \frac{I_{{ t } +}}{V_{{ t } +}} \right)
\end{eqnarray}
\end{subequations}
and represent a simpler alternative to \gl~(\ref{eq-tvs}) in cases of typical {\iv} asymmetries
(reasonably small $\gamma$), as often the case in experiment.\cite{Baldea:2019d,Baldea:2019h}

\subsection{Conclusion}
\label{sec:conclusion}

By deducing \gl~\ref{eq-approx}), in this paper I aimed at providing experimentalists with an extremely simple
recipe (``five-thirds protocol'')
that allows straightforward extraction of the microscopic parameters of molecular tunnel junctions obviating {\iv}
data fitting. As anticipated by the title, a pocket calculator is all what one needs for
the few arithmetic operations to be performed in applying \gl~\ref{eq-approx}).

By validating the ``five-thirds protocol'' for the platforms most commonly used to fabricate molecular tunnel junctions,
I aimed at convincing the molecular electronics community
on the generality of this approach enabling to gain important insight into the 
molecular structure-tunneling transport relationship. As seen, it can be applied even in situations
beyond the reach of the (by now) standard TVS-based approach \cite{Baldea:2012a}
whose broad usefulness has already be recognized.\cite{Frisbie:23}

\section*{Acknowledgments}
This research did not receive any specific financial support
but benefited from computational support by the
state of Baden-W\"urttemberg through bwHPC and the German Research Foundation through
Grant No.~INST 40/575-1 FUGG (bwUniCluster 2, bwForCluster/HELIX, and JUSTUS 2 cluster).
\balance

\renewcommand\refname{References}
\providecommand*{\mcitethebibliography}{\thebibliography}
\csname @ifundefined\endcsname{endmcitethebibliography}
{\let\endmcitethebibliography\endthebibliography}{}

% 
% \bibliography{ms.bib}

\begin{mcitethebibliography}{67}
\providecommand*{\natexlab}[1]{#1}
\providecommand*{\mciteSetBstSublistMode}[1]{}
\providecommand*{\mciteSetBstMaxWidthForm}[2]{}
\providecommand*{\mciteBstWouldAddEndPuncttrue}
  {\def\EndOfBibitem{\unskip.}}
\providecommand*{\mciteBstWouldAddEndPunctfalse}
  {\let\EndOfBibitem\relax}
\providecommand*{\mciteSetBstMidEndSepPunct}[3]{}
\providecommand*{\mciteSetBstSublistLabelBeginEnd}[3]{}
\providecommand*{\EndOfBibitem}{}
\mciteSetBstSublistMode{f}
\mciteSetBstMaxWidthForm{subitem}
{(\emph{\alph{mcitesubitemcount}})}
\mciteSetBstSublistLabelBeginEnd{\mcitemaxwidthsubitemform\space}
{\relax}{\relax}

\bibitem[Sze and Ng(2006)]{Sze:06}
S.~Sze and K.~K. Ng, in \emph{p-n Junctions}, John Wiley \& Sons, Ltd, 2006,
  pp. 77--133\relax
\mciteBstWouldAddEndPuncttrue
\mciteSetBstMidEndSepPunct{\mcitedefaultmidpunct}
{\mcitedefaultendpunct}{\mcitedefaultseppunct}\relax
\EndOfBibitem
\bibitem[Thompson(2014)]{Thompson:14}
M.~T. Thompson, in \emph{Intuitive Analog Circuit Design (Second Edition)}, ed.
  M.~T. Thompson, Newnes, Boston, Second Edition edn., 2014, pp. 53--86\relax
\mciteBstWouldAddEndPuncttrue
\mciteSetBstMidEndSepPunct{\mcitedefaultmidpunct}
{\mcitedefaultendpunct}{\mcitedefaultseppunct}\relax
\EndOfBibitem
\bibitem[Shockley(1949)]{Shockley:49}
W.~Shockley, \emph{Bell Syst. Tech. J.}, 1949, \textbf{28}, 435--489\relax
\mciteBstWouldAddEndPuncttrue
\mciteSetBstMidEndSepPunct{\mcitedefaultmidpunct}
{\mcitedefaultendpunct}{\mcitedefaultseppunct}\relax
\EndOfBibitem
\bibitem[Shockley(1950)]{Shockley:50}
W.~Shockley, \emph{Electrons and holes in semiconductors : with applications to
  transistor electronics}, Van Nostrand New York, New York, 1950\relax
\mciteBstWouldAddEndPuncttrue
\mciteSetBstMidEndSepPunct{\mcitedefaultmidpunct}
{\mcitedefaultendpunct}{\mcitedefaultseppunct}\relax
\EndOfBibitem
\bibitem[Sah \emph{et~al.}(1957)Sah, Noyce, and Shockley]{Sah:57}
C.-T. Sah, R.~N. Noyce and W.~Shockley, \emph{Proceedings of the IRE}, 1957,
  \textbf{45}, 1228--1243\relax
\mciteBstWouldAddEndPuncttrue
\mciteSetBstMidEndSepPunct{\mcitedefaultmidpunct}
{\mcitedefaultendpunct}{\mcitedefaultseppunct}\relax
\EndOfBibitem
\bibitem[Moll(1958)]{Moll:58}
J.~Moll, \emph{Proceedings of the IRE}, 1958, \textbf{46}, 1076--1082\relax
\mciteBstWouldAddEndPuncttrue
\mciteSetBstMidEndSepPunct{\mcitedefaultmidpunct}
{\mcitedefaultendpunct}{\mcitedefaultseppunct}\relax
\EndOfBibitem
\bibitem[Newns(1969)]{Newns:69b}
D.~M. Newns, \emph{Phys. Rev.}, 1969, \textbf{178}, 1123--1135\relax
\mciteBstWouldAddEndPuncttrue
\mciteSetBstMidEndSepPunct{\mcitedefaultmidpunct}
{\mcitedefaultendpunct}{\mcitedefaultseppunct}\relax
\EndOfBibitem
\bibitem[Anderson(1961)]{Anderson:61}
P.~W. Anderson, \emph{Phys. Rev.}, 1961, \textbf{124}, 41--53\relax
\mciteBstWouldAddEndPuncttrue
\mciteSetBstMidEndSepPunct{\mcitedefaultmidpunct}
{\mcitedefaultendpunct}{\mcitedefaultseppunct}\relax
\EndOfBibitem
\bibitem[Schmickler(1986)]{Schmickler:86}
W.~Schmickler, \emph{J. Electroanal. Chem.}, 1986, \textbf{204}, 31 -- 43\relax
\mciteBstWouldAddEndPuncttrue
\mciteSetBstMidEndSepPunct{\mcitedefaultmidpunct}
{\mcitedefaultendpunct}{\mcitedefaultseppunct}\relax
\EndOfBibitem
\bibitem[B\^aldea(2023)]{Baldea:2023a}
I.~B\^aldea, \emph{Phys. Chem. Chem. Phys.}, 2023, \textbf{25},
  19750--19763\relax
\mciteBstWouldAddEndPuncttrue
\mciteSetBstMidEndSepPunct{\mcitedefaultmidpunct}
{\mcitedefaultendpunct}{\mcitedefaultseppunct}\relax
\EndOfBibitem
\bibitem[B\^aldea(2024)]{Baldea:2024a}
I.~B\^aldea, \emph{Phys. Chem. Chem. Phys.}, 2024,  DOI
  10.1039/D3CP05046G\relax
\mciteBstWouldAddEndPuncttrue
\mciteSetBstMidEndSepPunct{\mcitedefaultmidpunct}
{\mcitedefaultendpunct}{\mcitedefaultseppunct}\relax
\EndOfBibitem
\bibitem[B\^aldea(2012)]{Baldea:2012a}
I.~B\^aldea, \emph{Phys. Rev. B}, 2012, \textbf{85}, 035442\relax
\mciteBstWouldAddEndPuncttrue
\mciteSetBstMidEndSepPunct{\mcitedefaultmidpunct}
{\mcitedefaultendpunct}{\mcitedefaultseppunct}\relax
\EndOfBibitem
\bibitem[Taherinia and Frisbie(2023)]{Frisbie:23}
D.~Taherinia and C.~D. Frisbie, \emph{Phys. Chem. Chem. Phys.}, 2023,
  \textbf{25}, 32305--32316\relax
\mciteBstWouldAddEndPuncttrue
\mciteSetBstMidEndSepPunct{\mcitedefaultmidpunct}
{\mcitedefaultendpunct}{\mcitedefaultseppunct}\relax
\EndOfBibitem
\bibitem[Xie \emph{et~al.}(2019)Xie, B\^aldea, and Frisbie]{Baldea:2019d}
Z.~Xie, I.~B\^aldea and C.~D. Frisbie, \emph{J. Am. Chem. Soc.}, 2019,
  \textbf{141}, 3670--3681\relax
\mciteBstWouldAddEndPuncttrue
\mciteSetBstMidEndSepPunct{\mcitedefaultmidpunct}
{\mcitedefaultendpunct}{\mcitedefaultseppunct}\relax
\EndOfBibitem
\bibitem[Xie \emph{et~al.}(2019)Xie, B\^aldea, and Frisbie]{Baldea:2019h}
Z.~Xie, I.~B\^aldea and C.~D. Frisbie, \emph{J. Am. Chem. Soc.}, 2019,
  \textbf{141}, 18182--18192\relax
\mciteBstWouldAddEndPuncttrue
\mciteSetBstMidEndSepPunct{\mcitedefaultmidpunct}
{\mcitedefaultendpunct}{\mcitedefaultseppunct}\relax
\EndOfBibitem
\bibitem[Nguyen \emph{et~al.}(2021)Nguyen, Xie, and Frisbie]{Frisbie:21a}
Q.~V. Nguyen, Z.~Xie and C.~D. Frisbie, \emph{J. Phys. Chem. C}, 2021,
  \textbf{125}, 4292--4298\relax
\mciteBstWouldAddEndPuncttrue
\mciteSetBstMidEndSepPunct{\mcitedefaultmidpunct}
{\mcitedefaultendpunct}{\mcitedefaultseppunct}\relax
\EndOfBibitem
\bibitem[Smaali \emph{et~al.}(2012)Smaali, Cl\'ement, Patriarche, and
  Vuillaume]{Vuillaume:12a}
K.~Smaali, N.~Cl\'ement, G.~Patriarche and D.~Vuillaume, \emph{ACS Nano}, 2012,
  \textbf{6}, 4639--4647\relax
\mciteBstWouldAddEndPuncttrue
\mciteSetBstMidEndSepPunct{\mcitedefaultmidpunct}
{\mcitedefaultendpunct}{\mcitedefaultseppunct}\relax
\EndOfBibitem
\bibitem[Tran \emph{et~al.}(2013)Tran, Smaali, Hardouin, Bricaud, O\c{c}afrain,
  Blanchard, Lenfant, Godey, Roncali, and Vuillaume]{Vuillaume:12c}
T.~K. Tran, K.~Smaali, M.~Hardouin, Q.~Bricaud, M.~O\c{c}afrain, P.~Blanchard,
  S.~Lenfant, S.~Godey, J.~Roncali and D.~Vuillaume, \emph{Adv. Mater.}, 2013,
  \textbf{25}, 427--431\relax
\mciteBstWouldAddEndPuncttrue
\mciteSetBstMidEndSepPunct{\mcitedefaultmidpunct}
{\mcitedefaultendpunct}{\mcitedefaultseppunct}\relax
\EndOfBibitem
\bibitem[Fracasso \emph{et~al.}(2013)Fracasso, Muglali, Rohwerder, Terfort, and
  Chiechi]{Fracasso:13}
D.~Fracasso, M.~I. Muglali, M.~Rohwerder, A.~Terfort and R.~C. Chiechi,
  \emph{J. Phys. Chem. C}, 2013, \textbf{117}, 11367--11376\relax
\mciteBstWouldAddEndPuncttrue
\mciteSetBstMidEndSepPunct{\mcitedefaultmidpunct}
{\mcitedefaultendpunct}{\mcitedefaultseppunct}\relax
\EndOfBibitem
\bibitem[Guo \emph{et~al.}(2013)Guo, Zhou, and Tao]{Tao:13}
S.~Guo, G.~Zhou and N.~Tao, \emph{Nano Lett.}, 2013, \textbf{13},
  4326--4332\relax
\mciteBstWouldAddEndPuncttrue
\mciteSetBstMidEndSepPunct{\mcitedefaultmidpunct}
{\mcitedefaultendpunct}{\mcitedefaultseppunct}\relax
\EndOfBibitem
\bibitem[Wu \emph{et~al.}(2013)Wu, Bai, Sanvito, and Hou]{Hou:13}
K.~Wu, M.~Bai, S.~Sanvito and S.~Hou, \emph{J. Chem. Phys.}, 2013,
  \textbf{139}, 194703\relax
\mciteBstWouldAddEndPuncttrue
\mciteSetBstMidEndSepPunct{\mcitedefaultmidpunct}
{\mcitedefaultendpunct}{\mcitedefaultseppunct}\relax
\EndOfBibitem
\bibitem[Lo \emph{et~al.}(2015)Lo, Bi, Li, Jung, and Yu]{Yu:15}
W.-Y. Lo, W.~Bi, L.~Li, I.~H. Jung and L.~Yu, \emph{Nano Lett.}, 2015,
  \textbf{15}, 958--962\relax
\mciteBstWouldAddEndPuncttrue
\mciteSetBstMidEndSepPunct{\mcitedefaultmidpunct}
{\mcitedefaultendpunct}{\mcitedefaultseppunct}\relax
\EndOfBibitem
\bibitem[Xiang \emph{et~al.}(2016)Xiang, Wang, Wang, Sun, Hou, and Liao]{Ho:15}
A.~Xiang, M.~Wang, H.~Wang, H.~Sun, S.~Hou and J.~Liao, \emph{Chem. Phys.},
  2016, \textbf{465-466}, 40--45\relax
\mciteBstWouldAddEndPuncttrue
\mciteSetBstMidEndSepPunct{\mcitedefaultmidpunct}
{\mcitedefaultendpunct}{\mcitedefaultseppunct}\relax
\EndOfBibitem
\bibitem[Nose \emph{et~al.}(2015)Nose, Dote, Sato, Yamamoto, Ishii, and
  Noguchi]{Yutaka:15}
D.~Nose, K.~Dote, T.~Sato, M.~Yamamoto, H.~Ishii and Y.~Noguchi, \emph{J. Phys.
  Chem. C}, 2015, \textbf{119}, 12765--12771\relax
\mciteBstWouldAddEndPuncttrue
\mciteSetBstMidEndSepPunct{\mcitedefaultmidpunct}
{\mcitedefaultendpunct}{\mcitedefaultseppunct}\relax
\EndOfBibitem
\bibitem[Kovalchuk \emph{et~al.}(2015)Kovalchuk, Abu-Husein, Fracasso, Egger,
  Zojer, Zharnikov, Terfort, and Chiechi]{Chiechi:15a}
A.~Kovalchuk, T.~Abu-Husein, D.~Fracasso, D.~Egger, E.~Zojer, M.~Zharnikov,
  A.~Terfort and R.~Chiechi, \emph{Chem. Sci.}, 2015, \textbf{7},
  781--787\relax
\mciteBstWouldAddEndPuncttrue
\mciteSetBstMidEndSepPunct{\mcitedefaultmidpunct}
{\mcitedefaultendpunct}{\mcitedefaultseppunct}\relax
\EndOfBibitem
\bibitem[Jia \emph{et~al.}(2016)Jia, Migliore, Xin, Huang, Wang, Yang, Wang,
  Chen, Wang, Feng, Liu, Zhang, Qu, Tian, Ratner, Xu, Nitzan, and Guo]{Guo:16a}
C.~Jia, A.~Migliore, N.~Xin, S.~Huang, J.~Wang, Q.~Yang, S.~Wang, H.~Chen,
  D.~Wang, B.~Feng, Z.~Liu, G.~Zhang, D.-H. Qu, H.~Tian, M.~A. Ratner, H.~Q.
  Xu, A.~Nitzan and X.~Guo, \emph{Science}, 2016, \textbf{352},
  1443--1445\relax
\mciteBstWouldAddEndPuncttrue
\mciteSetBstMidEndSepPunct{\mcitedefaultmidpunct}
{\mcitedefaultendpunct}{\mcitedefaultseppunct}\relax
\EndOfBibitem
\bibitem[Xiang \emph{et~al.}(2016)Xiang, Wang, Jia, Lee, and Guo]{Guo:16b}
D.~Xiang, X.~Wang, C.~Jia, T.~Lee and X.~Guo, \emph{Chem. Rev.}, 2016,
  \textbf{116}, 4318--4440\relax
\mciteBstWouldAddEndPuncttrue
\mciteSetBstMidEndSepPunct{\mcitedefaultmidpunct}
{\mcitedefaultendpunct}{\mcitedefaultseppunct}\relax
\EndOfBibitem
\bibitem[Wang \emph{et~al.}(2016)Wang, Liu, Xiang, Sun, Zhao, Sun, Mei, Wu,
  Liu, Guo, Li, and Lee]{Lee:16}
Q.~Wang, R.~Liu, D.~Xiang, M.~Sun, Z.~Zhao, L.~Sun, T.~Mei, P.~Wu, H.~Liu,
  X.~Guo, Z.-L. Li and T.~Lee, \emph{ACS Nano}, 2016, \textbf{10},
  9695--9702\relax
\mciteBstWouldAddEndPuncttrue
\mciteSetBstMidEndSepPunct{\mcitedefaultmidpunct}
{\mcitedefaultendpunct}{\mcitedefaultseppunct}\relax
\EndOfBibitem
\bibitem[Jeong \emph{et~al.}(2016)Jeong, Jang, Kim, Hwang, Kim, and
  Lee]{Lee:16a}
H.~Jeong, Y.~Jang, D.~Kim, W.-T. Hwang, J.-W. Kim and T.~Lee, \emph{J. Phys.
  Chem. C}, 2016, \textbf{120}, 3564--3572\relax
\mciteBstWouldAddEndPuncttrue
\mciteSetBstMidEndSepPunct{\mcitedefaultmidpunct}
{\mcitedefaultendpunct}{\mcitedefaultseppunct}\relax
\EndOfBibitem
\bibitem[Li \emph{et~al.}(2016)Li, Lo, Cai, Zhang, and Yu]{Yu:16b}
L.~Li, W.-Y. Lo, Z.~Cai, N.~Zhang and L.~Yu, \emph{Chem. Sci.}, 2016,
  \textbf{7}, 3137--3141\relax
\mciteBstWouldAddEndPuncttrue
\mciteSetBstMidEndSepPunct{\mcitedefaultmidpunct}
{\mcitedefaultendpunct}{\mcitedefaultseppunct}\relax
\EndOfBibitem
\bibitem[Cai \emph{et~al.}(2016)Cai, Lo, Zheng, Li, Zhang, Hu, and Yu]{Yu:16c}
Z.~Cai, W.-Y. Lo, T.~Zheng, L.~Li, N.~Zhang, Y.~Hu and L.~Yu, \emph{J. Am.
  Chem. Soc.}, 2016,  10630--10635\relax
\mciteBstWouldAddEndPuncttrue
\mciteSetBstMidEndSepPunct{\mcitedefaultmidpunct}
{\mcitedefaultendpunct}{\mcitedefaultseppunct}\relax
\EndOfBibitem
\bibitem[Lo \emph{et~al.}(2016)Lo, Zhang, Cai, Li, and Yu]{Yu:16d}
W.-Y. Lo, N.~Zhang, Z.~Cai, L.~Li and L.~Yu, \emph{Acc. Chem. Res.}, 2016,
  \textbf{49}, 1852--1863\relax
\mciteBstWouldAddEndPuncttrue
\mciteSetBstMidEndSepPunct{\mcitedefaultmidpunct}
{\mcitedefaultendpunct}{\mcitedefaultseppunct}\relax
\EndOfBibitem
\bibitem[Yi \emph{et~al.}(2017)Yi, Izarova, Stuckart, Guérin, Thomas, Lenfant,
  Vuillaume, van Leusen, Duchoň, Nemšák, Bourone, Schmitz, and
  Kögerler]{Lenfant:17}
X.~Yi, N.~V. Izarova, M.~Stuckart, D.~Guérin, L.~Thomas, S.~Lenfant,
  D.~Vuillaume, J.~van Leusen, T.~Duchoň, S.~Nemšák, S.~D.~M. Bourone,
  S.~Schmitz and P.~Kögerler, \emph{J. Am. Chem. Soc.}, 2017, \textbf{139},
  14501--14510\relax
\mciteBstWouldAddEndPuncttrue
\mciteSetBstMidEndSepPunct{\mcitedefaultmidpunct}
{\mcitedefaultendpunct}{\mcitedefaultseppunct}\relax
\EndOfBibitem
\bibitem[Cai \emph{et~al.}(2018)Cai, Zhang, Awais, Filatov, and Yu]{Yu:18}
Z.~Cai, N.~Zhang, M.~A. Awais, A.~S. Filatov and L.~Yu, \emph{Angew. Chem. Int.
  Ed.}, 2018, \textbf{57}, 6442--6448\relax
\mciteBstWouldAddEndPuncttrue
\mciteSetBstMidEndSepPunct{\mcitedefaultmidpunct}
{\mcitedefaultendpunct}{\mcitedefaultseppunct}\relax
\EndOfBibitem
\bibitem[Jeong and Song(2018)]{Song:18a}
I.~Jeong and H.~Song, \emph{Appl. Spectr. Rev.}, 2018, \textbf{53},
  246--263\relax
\mciteBstWouldAddEndPuncttrue
\mciteSetBstMidEndSepPunct{\mcitedefaultmidpunct}
{\mcitedefaultendpunct}{\mcitedefaultseppunct}\relax
\EndOfBibitem
\bibitem[Valianti \emph{et~al.}(2019)Valianti, Cuevas, and
  Skourtis]{Cuevas:19a}
S.~Valianti, J.-C. Cuevas and S.~S. Skourtis, \emph{J. Phys. Chem. C}, 2019,
  \textbf{123}, 5907--5922\relax
\mciteBstWouldAddEndPuncttrue
\mciteSetBstMidEndSepPunct{\mcitedefaultmidpunct}
{\mcitedefaultendpunct}{\mcitedefaultseppunct}\relax
\EndOfBibitem
\bibitem[Gu \emph{et~al.}(2021)Gu, Peng, Chen, and Chen]{Gu:21}
M.-W. Gu, H.~H. Peng, I.-W.~P. Chen and C.-h. Chen, \emph{Nat. Mater.}, 2021,
  \textbf{20}, 658--664\relax
\mciteBstWouldAddEndPuncttrue
\mciteSetBstMidEndSepPunct{\mcitedefaultmidpunct}
{\mcitedefaultendpunct}{\mcitedefaultseppunct}\relax
\EndOfBibitem
\bibitem[Liu \emph{et~al.}(2021)Liu, Qiu, Soni, and Chiechi]{Chiechi:21}
Y.~Liu, X.~Qiu, S.~Soni and R.~C. Chiechi, \emph{Chem. Phys. Rev.}, 2021,
  \textbf{2}, 021303\relax
\mciteBstWouldAddEndPuncttrue
\mciteSetBstMidEndSepPunct{\mcitedefaultmidpunct}
{\mcitedefaultendpunct}{\mcitedefaultseppunct}\relax
\EndOfBibitem
\bibitem[Kim \emph{et~al.}(2022)Kim, Im, and Song]{Song:22a}
Y.~Kim, K.~Im and H.~Song, \emph{Materials}, 2022, \textbf{15}, 774\relax
\mciteBstWouldAddEndPuncttrue
\mciteSetBstMidEndSepPunct{\mcitedefaultmidpunct}
{\mcitedefaultendpunct}{\mcitedefaultseppunct}\relax
\EndOfBibitem
\bibitem[Carlotti \emph{et~al.}(2022)Carlotti, Soni, Kovalchuk, Kumar, Hofmann,
  and Chiechi]{Chiechi:22}
M.~Carlotti, S.~Soni, A.~Kovalchuk, S.~Kumar, S.~Hofmann and R.~C. Chiechi,
  \emph{ACS Phys. Chem. Au}, 2022, \textbf{2}, 179--190\relax
\mciteBstWouldAddEndPuncttrue
\mciteSetBstMidEndSepPunct{\mcitedefaultmidpunct}
{\mcitedefaultendpunct}{\mcitedefaultseppunct}\relax
\EndOfBibitem
\bibitem[Jang \emph{et~al.}(2023)Jang, He, and Yoon]{Jang:23}
J.~Jang, P.~He and H.~J. Yoon, \emph{Acc. Chem. Res.}, 2023, \textbf{56},
  1613--1622\relax
\mciteBstWouldAddEndPuncttrue
\mciteSetBstMidEndSepPunct{\mcitedefaultmidpunct}
{\mcitedefaultendpunct}{\mcitedefaultseppunct}\relax
\EndOfBibitem
\bibitem[B\^aldea(2024)]{Baldea:2024c}
I.~B\^aldea, \emph{Phys. Chem. Chem. Phys.}, 2024,  DOI
  10.1039/D2CP05110A\relax
\mciteBstWouldAddEndPuncttrue
\mciteSetBstMidEndSepPunct{\mcitedefaultmidpunct}
{\mcitedefaultendpunct}{\mcitedefaultseppunct}\relax
\EndOfBibitem
\bibitem[Schmickler and Tao(1997)]{Schmickler:97}
W.~Schmickler and N.~Tao, \emph{Electrochimica Acta}, 1997, \textbf{42}, 2809
  -- 2815\relax
\mciteBstWouldAddEndPuncttrue
\mciteSetBstMidEndSepPunct{\mcitedefaultmidpunct}
{\mcitedefaultendpunct}{\mcitedefaultseppunct}\relax
\EndOfBibitem
\bibitem[Han \emph{et~al.}(1997)Han, Durantini, Moore, Moore, Gust, Rez,
  Leatherman, Seely, Tao, and Lindsay]{Tao:97}
W.~Han, E.~N. Durantini, T.~A. Moore, A.~L. Moore, D.~Gust, P.~Rez,
  G.~Leatherman, G.~R. Seely, N.~Tao and S.~M. Lindsay, \emph{J. Phys. Chem.
  B}, 1997, \textbf{101}, 10719--10725\relax
\mciteBstWouldAddEndPuncttrue
\mciteSetBstMidEndSepPunct{\mcitedefaultmidpunct}
{\mcitedefaultendpunct}{\mcitedefaultseppunct}\relax
\EndOfBibitem
\bibitem[Wang \emph{et~al.}(2003)Wang, Lee, and Reed]{Reed:03}
W.~Wang, T.~Lee and M.~A. Reed, \emph{Phys. Rev. B}, 2003, \textbf{68},
  035416\relax
\mciteBstWouldAddEndPuncttrue
\mciteSetBstMidEndSepPunct{\mcitedefaultmidpunct}
{\mcitedefaultendpunct}{\mcitedefaultseppunct}\relax
\EndOfBibitem
\bibitem[Pobelov \emph{et~al.}(2008)Pobelov, Li, and Wandlowski]{Wandlowski:08}
I.~V. Pobelov, Z.~Li and T.~Wandlowski, \emph{J. Am. Chem. Soc.}, 2008,
  \textbf{130}, 16045-- 16054\relax
\mciteBstWouldAddEndPuncttrue
\mciteSetBstMidEndSepPunct{\mcitedefaultmidpunct}
{\mcitedefaultendpunct}{\mcitedefaultseppunct}\relax
\EndOfBibitem
\bibitem[Garrigues \emph{et~al.}(2016)Garrigues, Yuan, Wang, Mucciolo, Thompon,
  del Barco, and Nijhuis]{Nijhuis:16a}
A.~R. Garrigues, L.~Yuan, L.~Wang, E.~R. Mucciolo, D.~Thompon, E.~del Barco and
  C.~A. Nijhuis, \emph{Sci. Rep.}, 2016, \textbf{6}, 26517\relax
\mciteBstWouldAddEndPuncttrue
\mciteSetBstMidEndSepPunct{\mcitedefaultmidpunct}
{\mcitedefaultendpunct}{\mcitedefaultseppunct}\relax
\EndOfBibitem
\bibitem[Delmas \emph{et~al.}(2020)Delmas, Diez-Cabanes, van Dyck, Scheer,
  Costuas, and Cornil]{Cornil:20}
V.~Delmas, V.~Diez-Cabanes, C.~van Dyck, E.~Scheer, K.~Costuas and J.~Cornil,
  \emph{Phys. Chem. Chem. Phys.}, 2020, \textbf{22}, 26702--26706\relax
\mciteBstWouldAddEndPuncttrue
\mciteSetBstMidEndSepPunct{\mcitedefaultmidpunct}
{\mcitedefaultendpunct}{\mcitedefaultseppunct}\relax
\EndOfBibitem
\bibitem[Im \emph{et~al.}(2022)Im, Seo, and Song]{Song:22b}
K.~Im, D.-H. Seo and H.~Song, \emph{Crystals}, 2022, \textbf{12}, 767\relax
\mciteBstWouldAddEndPuncttrue
\mciteSetBstMidEndSepPunct{\mcitedefaultmidpunct}
{\mcitedefaultendpunct}{\mcitedefaultseppunct}\relax
\EndOfBibitem
\bibitem[Beebe \emph{et~al.}(2006)Beebe, Kim, Gadzuk, Frisbie, and
  Kushmerick]{Beebe:06}
J.~M. Beebe, B.~Kim, J.~W. Gadzuk, C.~D. Frisbie and J.~G. Kushmerick,
  \emph{Phys. Rev. Lett.}, 2006, \textbf{97}, 026801\relax
\mciteBstWouldAddEndPuncttrue
\mciteSetBstMidEndSepPunct{\mcitedefaultmidpunct}
{\mcitedefaultendpunct}{\mcitedefaultseppunct}\relax
\EndOfBibitem
\bibitem[B\^aldea(2015)]{Baldea:2015a}
I.~B\^aldea, \emph{Phys. Chem. Chem. Phys.}, 2015, \textbf{17},
  15756--15763\relax
\mciteBstWouldAddEndPuncttrue
\mciteSetBstMidEndSepPunct{\mcitedefaultmidpunct}
{\mcitedefaultendpunct}{\mcitedefaultseppunct}\relax
\EndOfBibitem
\bibitem[B\^aldea \emph{et~al.}(2015)B\^aldea, Xie, and Frisbie]{Baldea:2015b}
I.~B\^aldea, Z.~Xie and C.~D. Frisbie, \emph{Nanoscale}, 2015, \textbf{7},
  10465--10471\relax
\mciteBstWouldAddEndPuncttrue
\mciteSetBstMidEndSepPunct{\mcitedefaultmidpunct}
{\mcitedefaultendpunct}{\mcitedefaultseppunct}\relax
\EndOfBibitem
\bibitem[B\^aldea(2012)]{Baldea:2012e}
I.~B\^aldea, \emph{Europhys. Lett.}, 2012, \textbf{98}, 17010\relax
\mciteBstWouldAddEndPuncttrue
\mciteSetBstMidEndSepPunct{\mcitedefaultmidpunct}
{\mcitedefaultendpunct}{\mcitedefaultseppunct}\relax
\EndOfBibitem
\bibitem[Metzger(2015)]{Metzger:15}
R.~M. Metzger, \emph{Chem. Rev.}, 2015, \textbf{115}, 5056--5115\relax
\mciteBstWouldAddEndPuncttrue
\mciteSetBstMidEndSepPunct{\mcitedefaultmidpunct}
{\mcitedefaultendpunct}{\mcitedefaultseppunct}\relax
\EndOfBibitem
\bibitem[Johnson \emph{et~al.}(2016)Johnson, Wickramasinghe, Verani, and
  Metzger]{Metzger:16a}
M.~S. Johnson, L.~D. Wickramasinghe, C.~N. Verani and R.~M. Metzger, \emph{J.
  Phys. Chem. C}, 2016, \textbf{120}, 10578--10583\relax
\mciteBstWouldAddEndPuncttrue
\mciteSetBstMidEndSepPunct{\mcitedefaultmidpunct}
{\mcitedefaultendpunct}{\mcitedefaultseppunct}\relax
\EndOfBibitem
\bibitem[Xie \emph{et~al.}(2021)Xie, B\^aldea, Nguyen, and
  Frisbie]{Baldea:2021d}
Z.~Xie, I.~B\^aldea, Q.~Nguyen and C.~D. Frisbie, \emph{Nanoscale}, 2021,
  \textbf{13}, 16755 -- 16768\relax
\mciteBstWouldAddEndPuncttrue
\mciteSetBstMidEndSepPunct{\mcitedefaultmidpunct}
{\mcitedefaultendpunct}{\mcitedefaultseppunct}\relax
\EndOfBibitem
\bibitem[Sullivan \emph{et~al.}(2023)Sullivan, Morningstar, Castellanos-Trejo,
  Welker, and Jurchescu]{Jurchescu:23}
R.~P. Sullivan, J.~T. Morningstar, E.~Castellanos-Trejo, M.~E. Welker and O.~D.
  Jurchescu, \emph{Nano Lett.}, 2023, \textbf{23}, 10864--10870\relax
\mciteBstWouldAddEndPuncttrue
\mciteSetBstMidEndSepPunct{\mcitedefaultmidpunct}
{\mcitedefaultendpunct}{\mcitedefaultseppunct}\relax
\EndOfBibitem
\bibitem[Metzger(2018)]{Metzger:18}
R.~Metzger, \emph{Nanoscale}, 2018, \textbf{10}, 10316--10332\relax
\mciteBstWouldAddEndPuncttrue
\mciteSetBstMidEndSepPunct{\mcitedefaultmidpunct}
{\mcitedefaultendpunct}{\mcitedefaultseppunct}\relax
\EndOfBibitem
\bibitem[Gupta \emph{et~al.}(2023)Gupta, Fereiro, Bayat, Pritam, Zharnikov, and
  Mondal]{Mondal:23}
R.~Gupta, J.~A. Fereiro, A.~Bayat, A.~Pritam, M.~Zharnikov and P.~C. Mondal,
  \emph{Nat. Rev. Chem.}, 2023, \textbf{7}, 106--122\relax
\mciteBstWouldAddEndPuncttrue
\mciteSetBstMidEndSepPunct{\mcitedefaultmidpunct}
{\mcitedefaultendpunct}{\mcitedefaultseppunct}\relax
\EndOfBibitem
\bibitem[Zotti \emph{et~al.}(2010)Zotti, Kirchner, Cuevas, Pauly, Huhn, Scheer,
  and Erbe]{Zotti:10}
L.~A. Zotti, T.~Kirchner, J.-C. Cuevas, F.~Pauly, T.~Huhn, E.~Scheer and
  A.~Erbe, \emph{Small}, 2010, \textbf{6}, 1529--1535\relax
\mciteBstWouldAddEndPuncttrue
\mciteSetBstMidEndSepPunct{\mcitedefaultmidpunct}
{\mcitedefaultendpunct}{\mcitedefaultseppunct}\relax
\EndOfBibitem
\bibitem[Widawsky \emph{et~al.}(2009)Widawsky, Kamenetska, Klare, Nuckolls,
  Steigerwald, Hybertsen, and Venkataraman]{Venkataraman:09a}
J.~R. Widawsky, M.~Kamenetska, J.~Klare, C.~Nuckolls, M.~L. Steigerwald, M.~S.
  Hybertsen and L.~Venkataraman, \emph{Nanotechnology}, 2009, \textbf{20},
  434009\relax
\mciteBstWouldAddEndPuncttrue
\mciteSetBstMidEndSepPunct{\mcitedefaultmidpunct}
{\mcitedefaultendpunct}{\mcitedefaultseppunct}\relax
\EndOfBibitem
\bibitem[Lee and Reddy(2011)]{Reddy:11}
W.~Lee and P.~Reddy, \emph{Nanotechnology}, 2011, \textbf{22}, 485703\relax
\mciteBstWouldAddEndPuncttrue
\mciteSetBstMidEndSepPunct{\mcitedefaultmidpunct}
{\mcitedefaultendpunct}{\mcitedefaultseppunct}\relax
\EndOfBibitem
\bibitem[Tan \emph{et~al.}(2010)Tan, Sadat, and Reddy]{Tan:10}
A.~Tan, S.~Sadat and P.~Reddy, \emph{Appl. Phys. Lett.}, 2010, \textbf{96},
  013110\relax
\mciteBstWouldAddEndPuncttrue
\mciteSetBstMidEndSepPunct{\mcitedefaultmidpunct}
{\mcitedefaultendpunct}{\mcitedefaultseppunct}\relax
\EndOfBibitem
\bibitem[Xie \emph{et~al.}(2015)Xie, B\^aldea, Smith, Wu, and
  Frisbie]{Baldea:2015d}
Z.~Xie, I.~B\^aldea, C.~Smith, Y.~Wu and C.~D. Frisbie, \emph{ACS Nano}, 2015,
  \textbf{9}, 8022--8036\relax
\mciteBstWouldAddEndPuncttrue
\mciteSetBstMidEndSepPunct{\mcitedefaultmidpunct}
{\mcitedefaultendpunct}{\mcitedefaultseppunct}\relax
\EndOfBibitem
\bibitem[Chen \emph{et~al.}(2024)Chen, B\^aldea, Yu, Liang, Li, Koren, and
  Xie]{Baldea:2024d}
Y.~Chen, I.~B\^aldea, Y.~Yu, Z.~Liang, M.-D. Li, E.~Koren and Z.~Xie,
  \emph{Langmuir}, 2024,  DOI 10.1021/acs.langmuir.3c03759\relax
\mciteBstWouldAddEndPuncttrue
\mciteSetBstMidEndSepPunct{\mcitedefaultmidpunct}
{\mcitedefaultendpunct}{\mcitedefaultseppunct}\relax
\EndOfBibitem
\bibitem[Su \emph{et~al.}(2023)Su, Zhang, Pan, Liang, Wang, and Guo]{Su:23}
L.~Su, Y.~Zhang, Q.~Pan, H.~Liang, H.~Wang and C.~Guo, \emph{New J. Chem.},
  2023, \textbf{47}, 17277--17283\relax
\mciteBstWouldAddEndPuncttrue
\mciteSetBstMidEndSepPunct{\mcitedefaultmidpunct}
{\mcitedefaultendpunct}{\mcitedefaultseppunct}\relax
\EndOfBibitem
\bibitem[Sergani \emph{et~al.}(2013)Sergani, Furmansky, and
  Visoly-Fisher]{Sergani:13}
S.~Sergani, Y.~Furmansky and I.~Visoly-Fisher, \emph{Nanotechnology}, 2013,
  \textbf{24}, 455204\relax
\mciteBstWouldAddEndPuncttrue
\mciteSetBstMidEndSepPunct{\mcitedefaultmidpunct}
{\mcitedefaultendpunct}{\mcitedefaultseppunct}\relax
\EndOfBibitem
\end{mcitethebibliography}
% 
% 
\end{document}